\documentclass[10pt,%
  reprint,
  superscriptaddress,
 amsmath,amssymb,
 aps,
 pre,
]{revtex4-1}

\usepackage{graphicx}% Include figure files
\usepackage{dcolumn}% Align table columns on decimal point
\usepackage{bm}% bold math
\usepackage[colorlinks=true,allcolors=blue]{hyperref}
\usepackage{amsmath}
\usepackage{SIunits}
\usepackage{color}

\renewcommand{\d}{\mathrm{d}}
\newcommand{\D}{\mathrm{D}}
\renewcommand{\vec}[1]{\bm{#1}}
\renewcommand{\v}{{\mathrm{v}}}
\newcommand{\sgn}[1]{\mathrm{sgn}(#1)}

\newcommand{\shr}{\tilde{v}_{xx}^0}
\newcommand{\shrfree}{\tilde{v}_{xx}^\mathrm{free}}

\newcommand{\xb}{\bar{x}}

\renewcommand{\v}{{\mathrm{v}}}
\newcommand{\vb}{\bar{\mathrm{v}}}

\definecolor{mmcolor}{rgb}{0.8, 0.0, 0.2}

\definecolor{micolor}{rgb}{0.0, 0.7, 0.2}

\begin{document}

% \title{Stability of pure shear deformation in active matter with both scalar and polar fields}
\title{Deforming polar active matter in a scalar field gradient}

\author{Muhamet Ibrahimi}
\affiliation{%
 Aix Marseille Univ, Université de Toulon, CNRS, CPT (UMR 7332), Turing Centre for Living Systems, Marseille, France
}%
\author{Matthias Merkel}%
\email{matthias.merkel@cnrs.fr}
\affiliation{%
 Aix Marseille Univ, Université de Toulon, CNRS, CPT (UMR 7332), Turing Centre for Living Systems, Marseille, France
}%
\date{\today}

\begin{abstract}
Active matter with local polar or nematic order is subject to the well-known Simha-Ramaswamy instability.
It is so far unclear how, despite this instability, biological tissues can undergo robust active anisotropic deformation during animal morphogenesis.
Here we show that protein concentration gradients (e.g.\ morphogen gradients), which are known to control large-scale coordination among cells, can stabilize such deformations.
To this end, we study a hydrodynamic model of an active polar material. 
To account for the effect of the protein gradient, the polar field is coupled to the boundary-provided gradient of a scalar field that also advects with material flows.
Focusing on the large system size limit, we show in particular:
(i) The system can be stable for an effectively extensile coupling between scalar field gradient and active stresses, i.e.\ gradient-extensile coupling, while it is always unstable for a gradient-contractile coupling.
Intriguingly, there are many systems in the biological literature that are gradient-extensile, while we could not find any that are clearly gradient-contractile.
% This suggests that nature might just not have evolved a gradient-contractile coupling.
(ii) Stability is strongly affected by the way polarity magnitude is controlled.
% (iii) In a pure-shear deforming system, linear stability is described by ``co-deforming'' perturbation modes whose growth rate can change over time. This leads to parameter regimes where the system is stable up to the existence of transiently growing modes.
Taken together, our findings, if experimentally confirmed, suggest new developmental principles that are directly rooted in active matter physics.
\end{abstract}

\maketitle

\section{Introduction}
Active matter is driven out of equilibrium by local injection of mechanical energy, which leads to new properties as compared to inert matter.
For instance, active matter with local polar or nematic order is known to exhibit the well-known Simha-Ramaswamy instability \cite{Simha2002}. This instability can lead to a spontaneous onset of flows \cite{Voituriez2005} or an instability of the homogeneously deforming state \cite{Simha2002,Voituriez2006}, and it
has already been observed in several biological systems, including cytoskeletal gels \cite{Sanchez2012}, bacterial swarms \cite{Beer2019,Liu2021a}, and cell monolayers in vitro \cite{Duclos2018}.
However, it is so far unclear whether this instability appears also in vivo during animal morphogenesis, and if not, how it is avoided.

One key process during animal morphogenesis is anisotropic tissue deformation, i.e.\ pure shear deformation of developing tissue \cite{Wolpert2015,Tada2012,Sutherland2019}.
While such deformation can be driven from outside, it is often also driven by active anisotropic stresses generated within the tissue itself \cite{Bertet2004,Bosveld2012,Tao2019,Keller2020,Maroudas-Sacks2021,Sermeus2022,Etournay2015,Dye2021}.
To obtain reproducible active anisotropic deformation, rotational symmetry needs to be broken; i.e.\ there needs to be some kind of directional information encoded in the system, and biological tissues have several ways to do this.
For instance, cells in a tissue can possess a polarity, which is defined by an anisotropic distribution of certain polarity proteins within the cell. In developing tissues, such cell polarity often exhibits large-scale ordered patterns \cite{Zallen2007,Aigouy2010,Merkel2014}. In some systems cell polarity can also induce an anisotropic distribution of the motor protein myosin within the cells, and thus control active anisotropic stresses \cite{Bertet2004,Bosveld2012}.
% This suggests a description of such tissues as active polar (or nematic) materials.
Such active polar or nematic materials should be prone to the Simha-Ramaswamy instability, and it is so far unclear what prevents it during development.

Animal morphogenesis relies crucially on large-scale protein concentration patterns \cite{Wolpert2015}. Such proteins, called morphogens, are important for the long-range coordination among cells during morphogenesis.
In particular, in several tissues, morphogen gradients are also known to control the direction of cell polarity \cite{Zallen2004,Zallen2007,Bosveld2012,Merkel2014,Butler2017,Lavalou2021}.
% In some cases, cellular polarization is believed to appear in the presence of a protein gradient only. Examples are myosin polarity in the germ band of the fruit fly \cite{Zallen2004,Lavalou2021} and Dachs polarity in the wing and eye discs of the fly \cite{Brittle2012}.
% In other cases, like for Core/Frizzled polarity, cells are believed to polarize also in the absence of any long-range protein gradient \cite{Devenport2014}, while its direction can still be controlled by protein gradients \cite{Butler2017,Merkel2014}.
However, it is so far unknown if such protein gradients could help stabilize anisotropic tissue deformation.

\begin{figure*}
     \includegraphics[width = 12.6 cm]{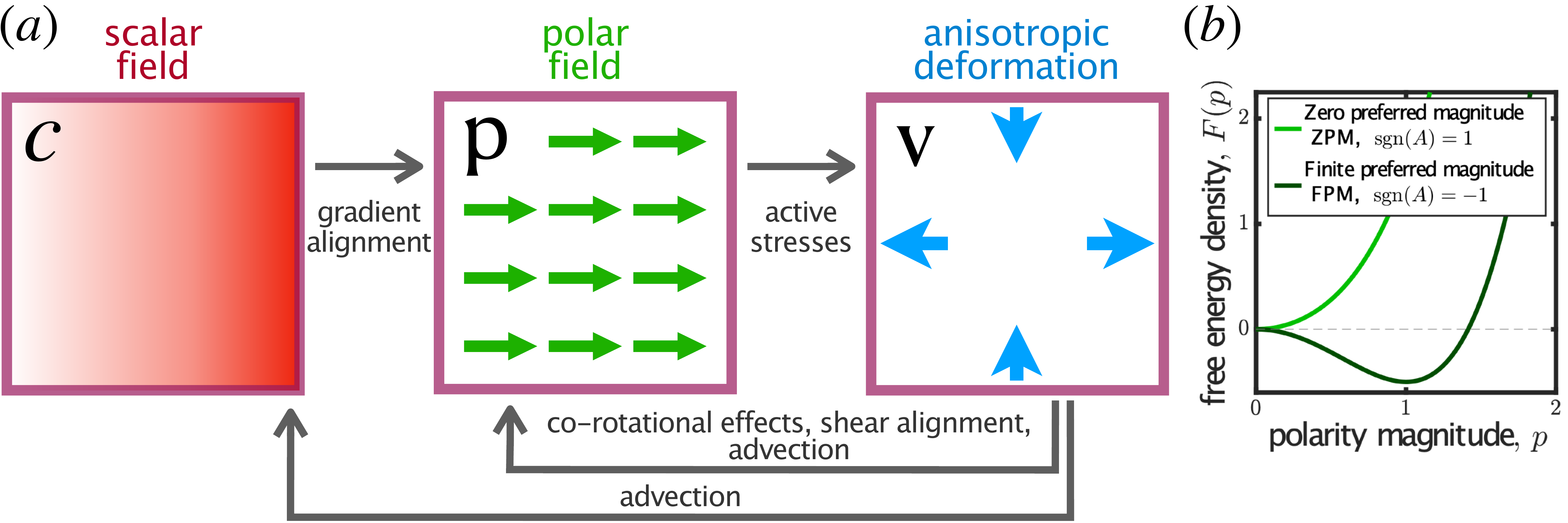}
    \caption{$(a)$ Schematic representation of the system: the gradient of a scalar field $c$ (left) directs a polar field $\vec{p}$ (center), which induces active anisotroic stresses, causing flows $\vec{\v}$ (right). Conversely, the flows affect the scalar and polar fields. $(b)$ Free energy densities for the two different polarity models discussed here, which describe polarity with zero and finite preferred magnitude, respectively.}
    \label{fig:three fields}
\end{figure*}

From a physics perspective, large-scale dynamics of biological tissues can be described using hydrodynamic active matter models \cite{Marchetti2013,Julicher2018}.
Such models correspond to expansions describing deviations away from a thermodynamic equilibrium state. They describe materials by averaging over their small-scale features, focusing on the dynamics on long length and time scales.
Indeed, in the recent past such hydrodynamic models have been successful in describing the multiple interactions between actively driven tissue deformation, cell polarity, and protein concentration fields \cite{Aigouy2010,Wartlick2011,Merkel2014,Banerjee2015,Etournay2015,Blanch-Mercader2017,Saw2017,Kawaguchi2017,Streichan2018,Duclos2018,Munster2019,Blanch-Mercader2021,Dye2021}.

Here we use this approach to study under which conditions a gradient of a protein that advects with tissue flows can help stabilize polarity-controlled anisotropic tissue deformation. 
We describe tissue flows by a velocity field $\vec{\v}$, cell polarity by a polarity field $\vec{p}$, and the protein concentration field by a scalar field $c$, where we impose a gradient via the boundary conditions. 
Model details are described in \autoref{sec:model}.
To examine this model, we first discuss the two limiting cases without polar field (\autoref{sec:scalar field only}) and without scalar field (\autoref{sec:polar field only}), before examining in detail the full model including both scalar and polar fields (\autoref{sec:scalar and polar field}).
Finally, we discuss under which conditions diffusion and local polarity alignment can stabilize systems of a finite size (\autoref{sec:system size effects}).

For the infinite-system-size limit, we find that gradient-contractile systems, i.e.\ systems with an effective contractile coupling between scalar field gradient and active stresses, are always unstable.  
Stable tissue deformation is only possible in gradient-extensile systems.
Intriguingly, up to one potential exception, effective gradient-extensile coupling is the only coupling that we could find in the biology literature across a multitude of multi-cellular animals.
We further show that the stability of deformation strongly depends on how the magnitude of cell polarity is controlled.
Taken together, our work suggests new potential developmental principles that are directly rooted in active matter physics.

\section{Model}
\label{sec:model}
% TS: We use continuum description of active matter to investigate the stability of homogeneous deformation in polar matter which is subject to active flows and an aligning scalar field.
% TS: We describe our material as a viscous incompressible fluid, given that deformation processes occur at the timescales typically from hours to days. However, this is a matter of debate...
\subsection{Bulk dynamics}
We study in 2D the interaction of a scalar field $c(\vec{r})$ describing a protein concentration field, a polar field $\vec{p}(\vec{r})$ describing cell polarity, and a velocity field $\vec{\v}(\vec{r})$ describing tissue flows (\autoref{fig:three fields}a). 

For the scalar field $c$ we focus on simple advective, diffusive dynamics :
\begin{equation} \label{eq:scalar field dynamics}
  \frac{\d c}{\d t} = D\partial_i^2 c,
\end{equation}
where we write the advective time derivative as $\d c/\d t = \partial c/\partial t + \v_i(\partial_i c)$, and $\partial_i$ denotes the partial derivative with respect to the spatial coordinate $r_i$. Here and in the following, we use Einstein notation, and we label spatial dimensions by Latin indices $i,j$.

To describe polarity dynamics, we introduce an effective free energy:
\begin{equation}\label{eq:free energy polarity}
\mathcal{F} = \int \left[ F(p) + \frac{K}{2}(\partial_j p_i)(\partial_j p_i) \right]\,\d^2r,
\end{equation} 
where the free energy density $F(p)$ controls the polarity magnitude $p=\vert\vec{p}\vert$, and the second term in the integrand controls local polarity alignment, where $K>0$. The latter is the Frank free energy in the one-constant approximation \cite{deGennes1995}. 
Unless stated otherwise, we will set:
\begin{equation}\label{eq:free energy polarity norm}
  F(p) = \frac{A}{2}p^2 + \frac{B}{4}p^4,
\end{equation}
where $B>0$ and $A$ can be either positive or negative.  The case $A<0$ corresponds to a polarity with a finite preferred magnitude (FPM), whereas $A>0$ corresponds to a polarity with a zero preferred magnitude (ZPM; \autoref{fig:three fields}b).

% We we can absorb $B>0$ by rescaling the polarity magnitude, and so we set $B=\vert A\vert$. (SMH rescaling later)
We use the following polarity dynamics \cite{Kruse2005,Julicher2018}:
\begin{equation} \label{eq:polarity}
\frac{\D p_i}{\D t} = \frac{1}{\gamma} h_i - \nu \tilde{v}_{ij}p_j + \beta\partial_i c.
\end{equation}
Here, the time derivative is a co-rotational derivative defined as $\D p_i/\D t = \partial p_i/\partial t + \v_j(\partial_j p_i)  + \Omega_{ij}p_j$ with the flow vorticity $\Omega_{ij} = (\partial_i\v_j-\partial_j\v_i)/2$. 
The first right-hand-side (rhs) term represents the relaxation of the polarity free energy  with $\gamma>0$ being a rotational viscosity and $h_i = -\frac{\delta\mathcal{F}}{\delta p_i}$ being the molecular field.
The second rhs term is a shear alignment term with coefficient $\nu$ and pure shear rate $\tilde{v}_{ij}=(\partial_i\v_j+\partial_j\v_i)/2$, where here and in the following, a tilde indicates the symmetric, traceless part of a tensor. For $\nu<0$, polarity tends to locally align with the extending direction of shear flow, while for $\nu>0$, polarity tends to locally align with the contracting direction of shear flow.
The last rhs term is a coupling to the gradient of the scalar field with coefficient $\beta$. Because the dynamics is invariant with respect to the transformation $(\vec{p}, \beta)\mapsto(-\vec{p}, -\beta)$, we set in the following without loss of generality $\beta>0$.

We study tissue flow that is governed by incompressible viscous dynamics. We define the stress tensor as
\begin{equation}
    \sigma_{ij} = 2\eta \tilde{v}_{ij} - \Pi\delta_{ij} + \sigma^p_{ij} + \tilde\sigma^a_{ij},\label{eq:stress}
\end{equation}
where $\eta$ is the shear viscosity, $\Pi$ is the hydrostatic pressure, the passive stress due to the polarity $\vec{p}$ is
\begin{equation}
    \sigma^p_{ij} = \frac{\nu}{2}(p_i h_j + p_j h_i) -\frac{1}{2}(p_i h_j - p_j h_i) ,
\end{equation}
and we use the following expression for the active stress:
    \begin{equation}
    \tilde\sigma^a_{ij} = \alpha \left( p_i p_j - \frac{p^2}{2}\delta_{ij} \right).
    % +\bar{\alpha}\left[(\partial_i c)(\partial_j c) - \frac{\delta_{ij}}{d}(\nabla c)^2\right]
\end{equation}
We use this stress tensor together with force balance and the incompressibility condition:
\begin{align}
  \partial_i\sigma_{ij} &= 0 \\
  \partial_i\v_i &= 0.
\end{align}
In most of the following, we focus on the infinite-system-size limit, neglecting diffusion $D$ and Frank coefficient $K$. We discuss the effect of finite $D$ and $K$ only in \autoref{sec:system size effects}.
Moreover, we neglect the passive stress $\sigma_{ij}^p$.  This corresponds to the limit of a small polarity free energy, $\mathcal{F}\rightarrow 0$, while keeping $\mathcal{F}/\gamma$ constant.  Otherwise, the passive stress $\sigma_{ij}^p$ can be absorbed into a redefinition of $\alpha$ and the hydrostatic pressure.

While more coupling terms between $c$, $\vec{p}$, and $\vec{\v}$ could be added to this model \cite{Marchetti2013,Julicher2018}, we focus here on the terms directly supported by experimental data on animal morphogenesis \cite{Zallen2004,Bertet2004,Aigouy2010,Bosveld2012,Merkel2014,Etournay2015,Streichan2018}.
Note that the role of the scalar field in this model is different from earlier work, where it represented a concentration of active agents or chemical fuel, and thus scaled the active stresses \cite{Marchetti2013,Giomi2014}.  Instead, here the scalar field acts as an aligning field for the polarity \cite{Simha2002,Voituriez2006}, while also being advected by tissue flows.

\subsection{Boundary conditions}
We study the dynamics in a rectangular periodic box of prescribed time-dependent dimensions $L_x(t)\times L_y(t)$. Because of incompressibility, $L_x(t)L_y(t)=\mathrm{const.}$, and prescribing $L_x(t)$ corresponds to prescribing the average shear rate tensor $\tilde{v}^0_{ij}(t)$ with:
\begin{equation}
  \tilde{v}^0_{xx}(t) = \frac{1}{L_x}\,\frac{\d L_x}{\d t}, \label{eq:avg shear rate}
\end{equation}
the other diagonal element is $\tilde{v}^0_{yy}=-\tilde{v}^0_{xx}$, and the off-diagonal elements vanish, $\tilde{v}^0_{xy}=\tilde{v}^0_{yx}=0$.

The boundary conditions are periodic for all fields, except for a modification for the scalar field $c$ at the vertical boundary. We set for all $y\in[0,L_y)$:
\begin{equation}
    c(0,y) = c(L_x,y) - c_b\label{eq:BCs c}
\end{equation}
with fixed $c_b$.
We introduce this modification to ensure that a linear profile $c=xc_b/L_x$ is stationary.

When prescribing not the box dimensions but an external stress anisotropy $\tilde\sigma_{ij}^\mathrm{ext}(t)$, Eq.~\eqref{eq:stress} implies that the system will shear with rate $\shr(t) = (\tilde{\sigma}_{xx}^\mathrm{ext}(t)-\langle\tilde\sigma^a_{xx}\rangle)/2\eta$. We use $\tilde\sigma_{ij}^\mathrm{ext}=\langle\tilde\sigma_{ij}\rangle$, where $\tilde\sigma_{ij}$ is the symmetric, traceless part of $\sigma_{ij}$, and $\langle\cdot\rangle$ is the spatial average over the system.
In the following, we will study stability for any constant $\shr(t)$. Two specific cases are: (i) a system with fixed size, $\shr=0$, and (ii) a freely deforming system, $\tilde\sigma_{ij}^\mathrm{ext}(t)=0$, which has the box deformation rate 
$\shr(t) = -\langle\tilde\sigma^a_{xx}\rangle/2\eta$.

\subsection{Dimensionless dynamics}
\label{sec:dimless}
We non-dimensionalize the dynamics, Eqs.~\eqref{eq:scalar field dynamics}--\eqref{eq:BCs c}, by choosing $c_b$ as unit for the scalar field, $\sqrt{\vert A\vert/B}$ as polarity unit, $L_x(0)$ as length scale, $\vert\alpha A\vert/B$ as stress scale, and $\eta B/\vert\alpha A\vert$ as time scale.  In the rest of this article, we use the accordingly rescaled dimensionless quantities. The dimensionless dynamical equations are
\begin{align}
  \frac{\d c}{\d t} &= D\partial_i^2c \label{eq:dimless c}\\
  \frac{\D p_i}{\D t} &= -\frac{g(p)}{\tau}p_i - \nu \tilde{v}_{ij}p_j + \beta\partial_i c + \kappa\partial_j^2p_i\label{eq:dimless p}\\
  0 &= \partial_i^2\v_j - \partial_j\Pi' + \sgn{\alpha}\,\partial_i(p_ip_j) \label{eq:dimless v}\\
  \partial_i\v_i &= 0. \label{eq:dimless incompressibility}
\end{align}
Here, we defined $g(p) = F'(p)/\vert A\vert p$ such that with our choice in Eq.~\eqref{eq:free energy polarity norm}:
\begin{equation}
  g(p) = \sgn{A} + p^2.
\end{equation}
Here we have introduced the sign function, $\sgn{A}:=A/\vert A\vert$.
We moreover introduced the dimensionless time scale $\tau = \gamma/\vert A\vert$ over which the polarity magnitude relaxes,
a polarity alignment coefficient $\kappa=K/\gamma$, and we set $\Pi'=\Pi+p^2/2$.
% Focusing on the infinite-system-size limit, we neglected the diffusion term in Eq.~\eqref{eq:dimless c} and the polarity alignment term in Eq.~\eqref{eq:dimless p}.

In dimensionless units, 
the modified boundary condition becomes $c(0,y) = c(L_x,y) - 1$,
the box deformation rate for the freely deforming system is $\shr = -\sgn\alpha\langle p_x^2-p_y^2\rangle/4$,
and we define the box shear $l_x(t) := L_x(t)/L_x(0) \equiv L_x(t)$.
% For constant $\shr(t)$, the box shear is $l_x(t)=e^{\shr t}$.

\begin{figure*}
    \includegraphics[width = 12.6cm]{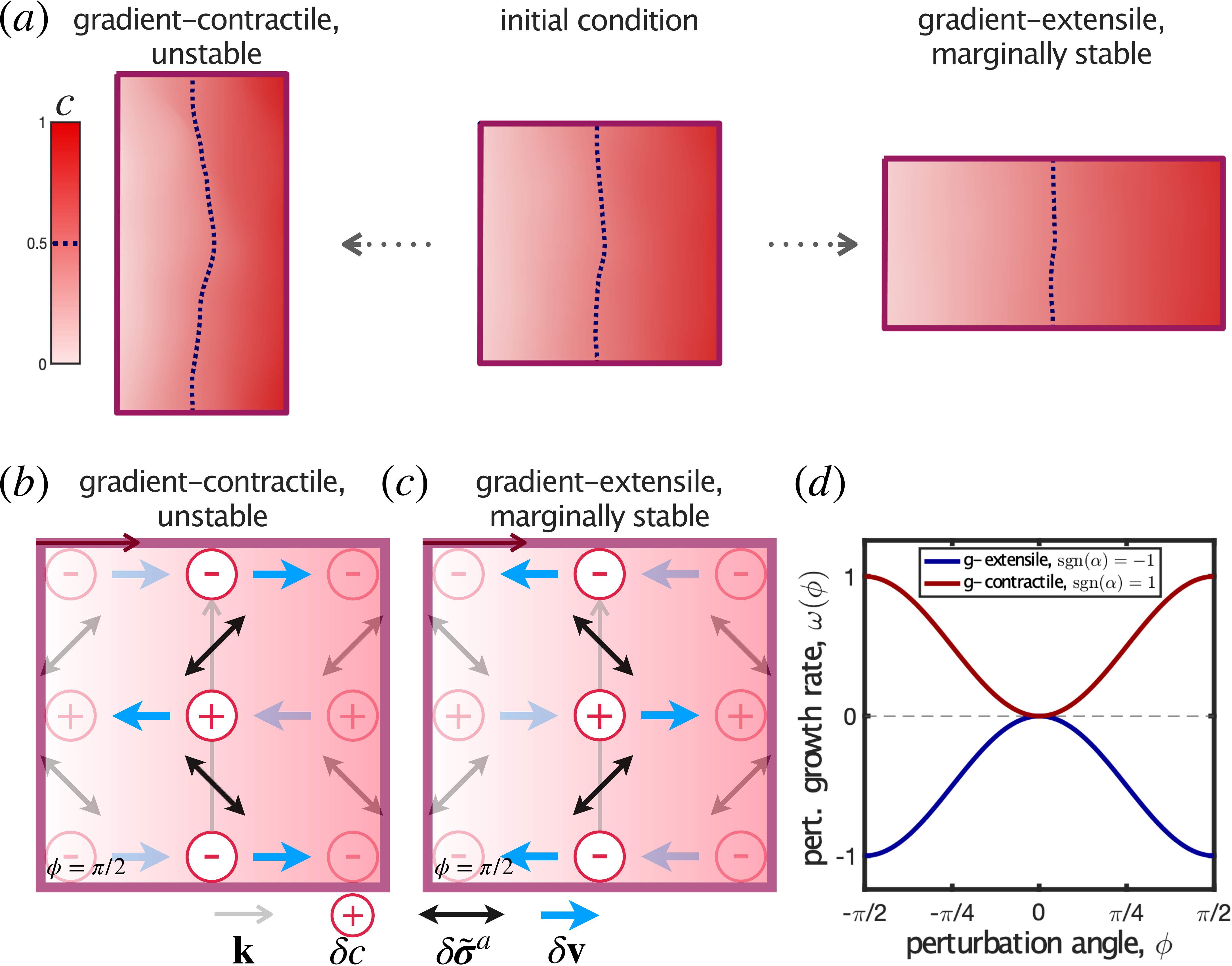}
    \caption{Linear stability for the scalar-only system, i.e.\ where $p_i = \partial_i c$. 
    $(a)$ Numerical solutions of the dynamics with a slightly perturbed initial state. The system is unstable for gradient-contractile coupling (left), and marginally stable for gradient-extensile coupling (right). The dashed dark line represents a $c$-isoline.
    $(b,c)$ Illustration of the linearized dynamics when the scalar field is perturbed by a mode with wave vector angle $\phi = \pi/2$. 
    $(b)$ unstable behavior for gradient-contractile coupling, $\sgn{\alpha} = 1$, and 
    $(c)$ stable behavior for gradient-extensile coupling, $\sgn{\alpha} = -1$. 
    $(d)$ Growth rate of a perturbation with given wave vector orientation $\phi$.
    }
     \label{fig:scalar field only}
\end{figure*}

\section{Scalar field only}
\label{sec:scalar field only}
We first discuss the special case without polar field.  This can also be regarded as the limit where the polarity relaxes to the scalar field gradient adibatically fast:
\begin{equation}
  p_i = \partial_i c.
\end{equation}
More precisely, this corresponds to the limit $\beta\rightarrow\infty$ while keeping $\beta\tau=1$, and using the polarity potential $F(p)=p^2/2$ (i.e.\ $g(p)=1$).
As a consequence, the active stress is given by the gradient of the scalar field $\tilde\sigma^a_{ij}=\sgn{\alpha}\,[(\partial_ic)(\partial_jc)-(\partial_kc)^2\delta_{ij}/2]$, like in Active Model H \cite{Tiribocchi2015a}.

% Note that with free boundary conditions, i.e.\ without externally applied stress, the system would deform with shear rate
% \begin{equation} 
%   \shrfree(t) = -\frac{\sgn{\alpha}}{4l_x^2(t)}. \label{eq:scalar free deformation rate}
% \end{equation}
Example numerical solutions of the dynamics for the freely deforming system are shown in \autoref{fig:scalar field only}a:  gradient-contractile systems, $\alpha>0$, are unstable, while gradient-extensile systems, $\alpha<0$, are marginally stable.

% In the following, we first discuss the linear stability of a fixed system size, $\shr=0$, before turning to deforming systems, $\shr\neq 0$.
% The flow field $\vec{\v}_0(\vec{r}, t)$ of this state is imposed by the boundary condition Eq.~\eqref{eq:avg shear rate}: 
% \begin{align}
%   \v_{0,i}(\vec{r}, t) &= \tilde{v}_{ij}^0(t)r_j.  \label{eq:scalar v0}
% \end{align}

% \mi{In \autoref{fig:scalar field only}a we show two different simulations of the system with a box width $l_x(0) = 1$ and a perturbed linear profile that is freely deforming  ($\tilde\sigma_{ij}^\mathrm{ext}(t)=0$). The system shows a typically unstable behavior in the gradient-contractile coupling (left) and a stable behavior in the gradient-extensile coupling (right). The $c = 0.5$ isolines in \autoref{fig:scalar field only}a reflect how the amplitude of perturbation evolves. Details of the simulations are explained in appendix~\ref{app: scalar only simulation}.}

\subsection{\texorpdfstring{Fixed system size, $\shr=0$}{Fixed system size}}
We first perform a linear stability analysis for a non-deforming system, $\shr=0$, by perturbing the scalar field $c$ around the linear profile:
\begin{align}
  c &= c_0 + \delta c,
\end{align}
% In the following, we will examine the stability of the homogeneously deforming state with a linear concentration profile $c_0(\vec{r}, t)$ with $\vec{r}=(x,y)$: 
where we defined the linear profile $c_0$ as:
\begin{align}
  c_0(\vec{r}, t) &= \frac{x}{l_x(t)} \label{eq:scalar_only_c0}
\end{align}
with $\vec{r}=(x,y)$.
We obtain for the growth rate $\omega$ of a perturbation with wave vector $\vec{k} = k(\cos{\phi},\sin{\phi})$ (\autoref{fig:scalar field only}d, appendix~\ref{app: scalar only}):
\begin{equation}
\omega(\vec{k}) := \frac{\partial_t\delta c(\vec{k}, t)}{\delta c(\vec{k}, t)} = \sgn{\alpha}\sin^2{\phi}.
\end{equation}
Thus, consistent with earlier work \cite{Kirkpatrick2019a}, the system is linearly unstable for gradient-contractile systems ($\alpha>0$), while it is marginally stable for gradient-extensile systems ($\alpha<0$), where the only modes that do not decay over time are those with wave vector angles $\phi=0$ and $\phi=\pi$.
The magnitudes of growth and decay rates can thus be up to $\pm 1$ in dimensionless units, which is four times the magnitude of the free deformation rate, which is $\shrfree = -\sgn{\alpha}/4$.

% SMH: maybe in discussuion:  Note that in this section (SMH check), ``contractile'' and ``extensile'' refer to the relative orientation between active stress and the \emph{gradient of the scalar field}.

Intuitively, the difference in behavior between gradient-extensile and gradient-contractile cases is illustrated in \autoref{fig:scalar field only}b,c, which represent the behavior of a perturbation with wave vector angles $\phi=\pi/2$. 
A perturbation $\delta c$ of the scalar field (red symbols in \autoref{fig:scalar field only}b,c) induces to first order a perturbation in the active stress nematic of $\delta\tilde\sigma^a_{xi}\sim k_i\delta c$, which have angles of $\pm\pi/4$ (black double arrows).
For the gradient-extensile case ($\alpha<0$, \autoref{fig:scalar field only}c), this stress perturbation generates flows (blue arrows) that advect regions with positive $\delta c$ in positive $x$ direction, leading to a local decrease in $c$ due to the overall $c$ gradient. Hence, for the gradient-extensile case this perturbation decays.  Analogously, for the gradient-contractile case, advection leads to an amplification of the $\pi/2$ mode (\autoref{fig:scalar field only}b).

% Intuitively, the difference in behavior depending on the sign of $\alpha$ can also be understood as created by a positive (gradient-extensile) or negative (gradient-contractile) interface tension between regions with low and high $c$, respectively \cite{Tiribocchi2015a}.

% For the contractile case, the system can only be stabilized through diffusion as long as only with wavelengths modes $k>k_\mathrm{min}$ are allowed (\autoref{fig:scalar field only}c).  In other words, a contractile system can only be stable if its dimension $L_y$ is smaller than a characteristic length scale
% \begin{equation}
%     L_c = \pi \sqrt{\frac{D}{\vert\shrfree\vert}}.
% \end{equation}
% This is analogous to the critical length scale beyond which the homogeneously deforming state in active polar and nematic matter becomes unstable \cite{Marchetti2013,Simha2002,Voituriez2005}.

\subsection{\texorpdfstring{Deforming system, $\shr\neq0$}{Deforming system}}
\label{sec:deforming system scalar field}
For a deforming system, the linear profile $c_0$ (Eq.~\eqref{eq:scalar_only_c0}) becomes distorted by the system's overall deformation, which means that it is not a stationary state with respect to the lab frame anymore. 
However, stationarity is still possible with respect to ``co-deforming'' coordinates $\vec{\bar r}=(\bar{x},\bar{y})$, which we define as (see appendix~\ref{app: co-deforming coordinates}):
\begin{align}
  \bar{x} &= \frac{x L_x(0)}{L_x(t)} \equiv l_x^{-1}(t)x \\
  \bar{y} &= \frac{y L_y(0)}{L_y(t)} \equiv l_x(t)y.
\end{align}
% Spatial partial derivatives of a quantity $q$ with respect to these coordinates change as $\bar\partial_x q := \partial_{\bar x} q(\bar{x}, \bar{y}, t) = L_x(t)\partial_x q(x, y, t)$ and $\bar\partial_y q := \partial_{\bar y} q(\bar{x}, \bar{y}, t) = L_x^{-1}(t)\partial_y q(x, y, t)$.  Finally, the new spatial coordinates also give rise to a new time derivative $\bar\partial_t q := \partial_{t} q(\bar{x}, \bar{y}, t)$.
These coordinates thus map any point $\bm{r}$ in the system at time $t$ to its affinely rescaled position $\bm{\bar{r}}$ at time zero.
As a consequence, the linear profile $c_0$ does not become distorted when written in co-deforming coordinates
\begin{equation}
  c_0(\vec{\bar r}, t) = \bar{x}.
\end{equation}
Hence, in co-deforming coordinates $c_0$ can be regarded as stationary. 

\begin{figure*}
     \centering
     \includegraphics[width = 12.6 cm]{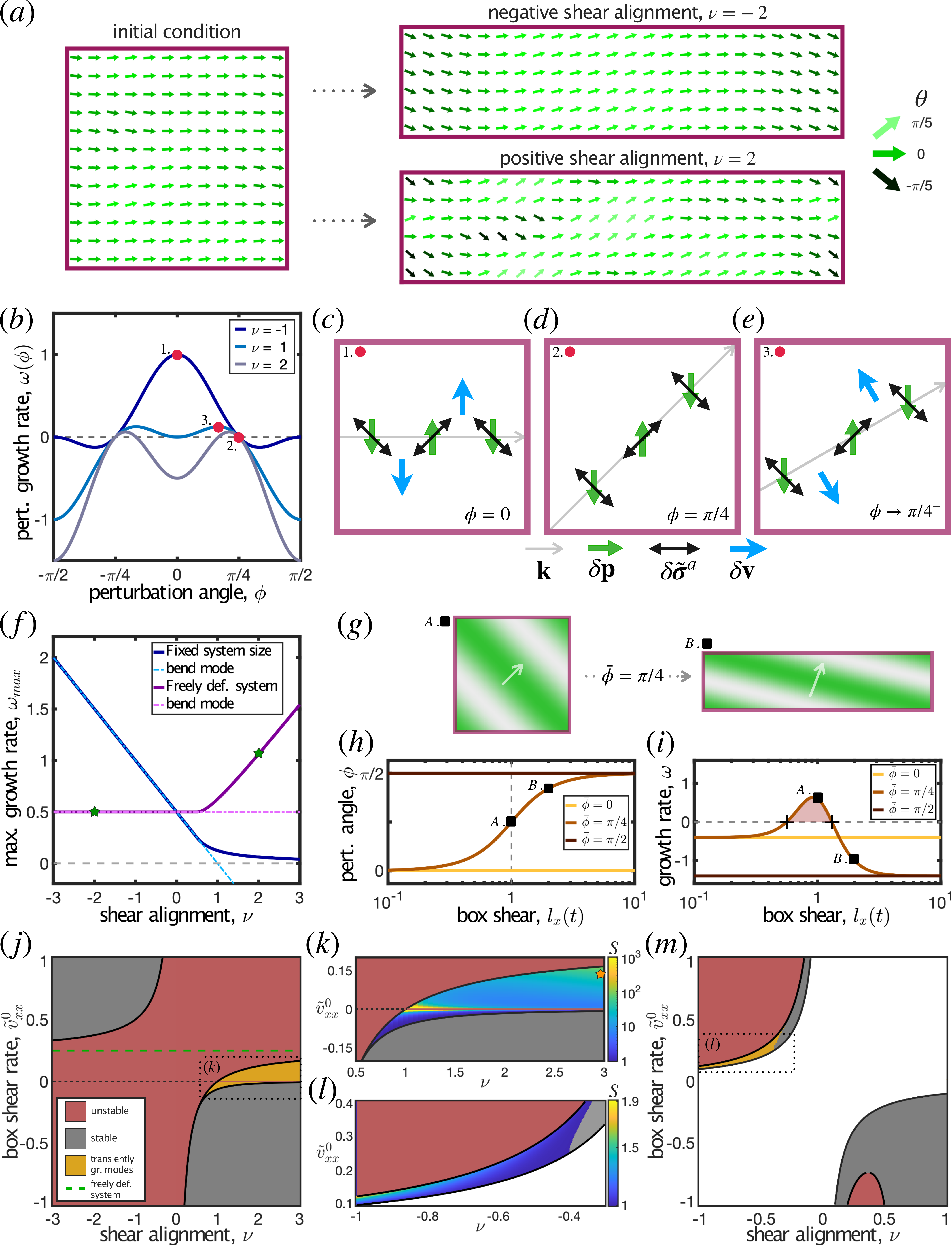}
     \caption{Linear stability of the polar-only system, $\beta=0$.
     $(a)$~Simulations of a freely deforming system with fixed-magnitude polarity and extensile active stresses, for two different values of the shear alignment parameter $\nu$.
     $(b)$~Growth rate $\omega$ of a perturbation with angle $\phi$ in a system with a fixed polarity magnitude (i.e.\ $\sgn{A} = -1$ and $\tau\rightarrow 0$) for three different $\nu$ values. 
     $(c)$-$(e)$~Schematics illustrating the perturbation dynamics for three different $\phi$ (\autoref{sec:polar field only - fixed}). 
     $(f)$~Perturbation growth rate maximized across all wave vectors, $\omega_{max}$, for fixed system (blue) and freely deforming system (purple). The dashed lines represent the growth rates of the respective bend modes ($\phi=0$). Green stars correspond to the $\nu$ values used for the simulations in panel a.
     $(g)$~Illustration of the change of the wave vector $\vec{k}$ measured in the lab frame (arrow) for a given co-deforming mode $\vec{\bar k}$ as the system is sheared from $l_x(t) = 1$ (A) to $l_x(t) = 2$ (B; compare Eqs.~\eqref{eq:kx_kbarx} and \eqref{eq:ky_kbary}). 
     $(h)$~Perturbation angle $\phi$ measured in the lab frame versus the box shear $l_x$, for three different co-deforming angles $\bar{\phi}$. 
     $(i)$~Change of the growth rates $\omega$ of three co-deforming perturbation modes (same as in panel h) as the box width $l_x$ changes during the box deformation. The co-deforming $\bar{\phi} = \pi/4$ mode transiently experiences a positive growth rate (positive $\omega$ values between `+' signs).
     Fixed polarity magnitude, $\shr = 1/8$, $\nu = 3$.
     $(j)$-$(m)$~Stability phase diagrams for $(j,k)$ fixed polarity magnitude and $(l,m)$ zero preferred polarity magnitude with $\tau=10$, where red, gray, and yellow respectively indicate unstable regions, stable regions, and stable regions with transiently growing modes. The green dashed line in $(j)$ represents the deformation rate of the freely deforming system. Black lines marking phase boundaries are analytical curves derived in appendix~\ref{app: polarity only}. $(k)$-$(l)$ Maximal perturbation amplification factor $S$ in the stable regions with transiently growing modes. Orange star in $(k)$ corresponds to the parameters used for panel i.}
     \label{fig:polar-only}
\end{figure*}

Co-deforming coordinates are useful also because they help us solve the linearized dynamics of the system, which is required to analyze the system's stability (appendix~\ref{app:linearized dynamics}). 
We find that the solutions of the linearized dynamics are ``co-deforming Fourier modes'':
\begin{equation}
  \delta c(\vec{\bar r}, t) = \int{\delta c(\vec{\bar k}, t)\,e^{i\vec{\bar k}\cdot\vec{\bar r}}\,\d^2\bar{k}}. \label{eq:codef fourier}
\end{equation}
We call $\vec{\bar k}$ a co-deforming wave vector.
% Note that we use the same symbol $\delta c$ for the real-space field and Fourier amplitudes; which of the two is meant will be indicated by the parameters.
A co-deforming Fourier mode with wave vector $\vec{\bar k}$ is distorted over time by the overall system deformation (\autoref{fig:polar-only}d, appendix~\ref{app:co-deforming coordinates definition}):
\begin{align}
    k_x(\vec{\bar k}, t) &= l_x^{-1}(t)\bar{k}_x \label{eq:kx_kbarx}\\
    k_y(\vec{\bar k}, t) &= l_x(t)\bar{k}_y.\label{eq:ky_kbary}
\end{align}
These relations will introduce a time dependence in the growth rate of co-deforming Fourier modes.

For the deforming scalar-only system, we obtain for the growth rate of a co-deforming Fourier mode with wave vector $\vec{\bar k}$ (see appendix~\ref{app: scalar only}):
\begin{equation}
  \omega(\vec{\bar k}, t) := \frac{\partial_t\delta c(\vec{\bar k}, t)}{\delta c(\vec{\bar k}, t)} = \sgn{\alpha}\frac{\sin^2{\phi(\vec{\bar k}, t)}}{l_x^2(t)}.\label{eq:lin dyn deforming}
\end{equation}
Here, the angle $\phi$ of the lab-frame wave vector $\vec{k}$ depends on $\vec{\bar k}$ and $t$ through Eqs.~\eqref{eq:kx_kbarx} and \eqref{eq:ky_kbary}.
As a consequence of the time-dependent right-hand side of Eq.~\eqref{eq:lin dyn deforming}, the solutions $\delta c(\vec{\bar k}, t)$ are generally not exponential in $t$ any more. 
However, we can still discuss the stability of the system.
Indeed, like in the case with fixed system size, the system is unstable in the gradient-contractile case, $\alpha>0$, while it is marginally stable in the gradient-extensile case, $\alpha<0$. 
% Hence, when a scalar field gradient controls active anisotropic stresses, a deforming viscous system is marginally stable only in the gradient-extensile case.
% or when stabilized by diffusion.
% - discuss

% - with given mode k bar
% - this is a mode whose wave vector change over time in real coordinates according to k(t)=
% - as a consequence, linearized dynamics contains time-dependent prefactors, where we find
% - ODE for delta c
% - bc of time dependence in ..., exponential in general no solution any more, but we can still discuss the stability of the system

% \subsection{Freely deforming system, $\shr=\shrfree$}
% - solution Lx(t) + plot
% - show plot amplitude over time of modes with bar angles 0, pi/6, pi/4, pi/3, pi/2 for freely deforming system compared to fixed system

\section{Polar field only}
\label{sec:polar field only}
We now revisit the limit where the scalar field plays no role, $\beta=0$, i.e.\ where our systems becomes a polar active matter system.  While such systems have been discussed before in the literature \cite{Marchetti2013,Simha2002,Voituriez2005}, we will shed light on a few new aspects while preparing for the section discussing the full model.

Example simulations of a freely deforming system with fixed polarity magnitude are shown in \autoref{fig:polar-only}a:  we find unstable behavior independent of the sign of the shear alignment $\nu$.

In our discussion below, we focus here on the extensile case, $\alpha<0$. For $\beta=0$ these results can be directly mapped to the contractile case.

% are analogous since the system dynamics  is invariant with respect to the mapping $(\alpha,\nu,\vec{p})\mapsto(-\alpha,-\nu,\bm{R}(\pi/2)\cdot\vec{p})$, where $\bm{R}(\pi/2)$ denotes a rotation by $\pi/2$.

% We separately discuss the two cases of finite preferred polarity magnitude, $\sgn{A}=-1$, and zero preferred polarity magnitude, $\sgn{A}=1$.

% , we report preliminary simulations of a polarity of fixed magnitude in a freely deforming system $\tilde\sigma_{ij}^\mathrm{ext}(t)=0$ (see appendix \ref{app: polarity only, fixed, simulations}). On an extensile coupling, the perturbed homogeneous state exhibits unstable behavior independently of the shear alignment parameter $\nu$. Note that for negative $\nu$ values the bend mode dominates the instability, and in a positive $\nu$ regime multiple modes are unstable.

% \mi{In the following, we continue with the linear stability analyses of the polar field only case in a systematic fashion.}

\subsection{\texorpdfstring{Fixed system size, $\shr=0$}{Fixed system size}}
\label{sec:polar field only - fixed}
Here we focus exclusively on the case of a finite preferred polarity magnitude (FPM, \autoref{fig:three fields}b). For fixed system size, $\shr=0$, there are no stationary solutions with finite polar order for zero preferred magnitude (ZPM).

% \subsubsection{Finite preferred polarity magnitude, $\sgn{A}=-1$}
We briefly revisit the stability of the polar ordered state \cite{Simha2002,Voituriez2005}:
\begin{align}
  \vec{p}_0 &= \hat{x}, \\
  \bm{\v}_{0} &= 0.
\end{align}
The system is unstable for every value of the shear alignment coefficient $\nu$ \cite{Simha2002}.
To prepare for the later sections, we now revisit the intuitive explanation of this for the limit of $\tau \rightarrow 0$, where the polarity magnitude is fixed to one, $\vert \vec{p}\vert = 1$  (\autoref{fig:polar-only}b,f, appendix~\ref{app: polarity only, fixed}). 
% In this limit, the growth rate of a polarity perturbation with wave vector $\vec{k}$ is (appendix~\ref{app: polarity only, fixed}) \cite{Simha2002}: 
% \begin{equation}
%     \omega(\vec{k}) = -\frac{1}{2}\cos{(2\phi)}\big[\nu \cos{(2\phi)}-1\big].  \label{eq:growth rate, polarity only, no deformation}
% \end{equation}
% This system is unstable for every value of the shear alignment coefficient $\nu$ (\autoref{fig:polar-only}c). 

For $\nu<1$ the bend mode $\phi=0$ is unstable (\autoref{fig:polar-only}b and blue dashed line in \autoref{fig:polar-only}f) \cite{Voituriez2005}.  The mechanism driving this instability is illustrated in \autoref{fig:polar-only}c:
For fixed polarity norm, perturbations in the polarity are exclusively orientational, i.e.\ $\vec{\delta p}$ is oriented along the $y$ axis (green arrows). 
This affects the linear perturbation in the active stress tensor $\delta\tilde{\sigma}^a_{ij}$, which is in general given by:
\begin{equation}
  \delta\tilde{\sigma}^a_{xi} = \alpha p_0\delta p_i.\label{eq:active stress-polarity perturbation}
\end{equation}
For $\vec{\delta p}$ oriented along the $y$ axis, Eq.~\eqref{eq:active stress-polarity perturbation} implies that $\delta\tilde{\sigma}^a_{ij}$ is oriented at angles of $\pm\pi/4$ with respect to the $x$ axis (black arrows). This active stress perturbation creates flow (blue arrows), which for $\nu<1$ acts through a combination of co-rotational and shear alignment effects to amplify the perturbation $\delta p_y$.

For $\nu\geq1$, where the bend mode is stable, there are always other unstable modes \cite{Simha2002}.
To see this, we first discuss the mode with $\phi=\pi/4$ (\autoref{fig:polar-only}d). In this case, the angle of the wave vector $\vec{k}$ is parallel or perpendicular to the angle of the active stress perturbation $\delta\tilde{\sigma}^a_{ij}$, and thus no flow appears due to incompressibility.
However, when slightly decreasing $\phi$ below $\pi/4$ (\autoref{fig:polar-only}e), flow appears, which acts through the co-rotational effect to amplify $\delta p_y$.
This co-rotational effect always dominates over the shear alignment when approaching $\phi=\pi/4$. 
% This is because at $\phi=\pi/4$, the direction of local pure shear is either perpendicular or parallel to the polarity direction $\vec{\hat{x}}$, leading to a zero shear alignment effect even for finite shear rates. 
As a consequence, there are always perturbations with angles $\phi$ below $\pi/4$ whose growth rate are positive.
% In other words, the co-rotational term will always dominate close to $\phi=\pi/4$, and lead for any value of $\nu$ to an amplification of the polarity perturbation for some angles $\phi$ right below $\pi/4$ (\autoref{fig:polar-only}a,b).

These results do not fundamentally change when allowing for a finite value of the polarity magnitude relaxation time, $\tau>0$, i.e.\ a ``soft'' polarity magnitude.
For $\nu<1$ the fastest growing mode is still the bend mode, whose growth rate is unaffected by $\tau$.
For $\nu>1$, the growth rate of the fastest growing mode decreases with increasing $\tau$. However  it is still positive for any $\tau$ (\autoref{fig: app fig 1}). 
Hence, the system with fixed size is always unstable, also for finite $\tau$.

\subsection{\texorpdfstring{Deforming system, $\shr\neq0$}{Deforming system}}
For a deforming system with given box shear rate $\shr$ the stationary, homogeneously deforming state is:
\begin{align}
  \vec{p}_0 &= p_0\hat{x}, \\
  \v_{0,i} &= \tilde{v}_{ij}^0r_j,
\end{align}
where the value of the stationary polarity magnitude $p_0$ depends on the given box shear rate $\shr$. 
In this section we discuss the stability of this stationary state using co-deforming perturbations (\autoref{sec:deforming system scalar field}, appendix~\ref{app: co-deforming coordinates}).

\subsubsection{Finite preferred polarity magnitude (FPM)}
% For finite preferred polarity magnitude, $\sgn{A}=-1$, the homogeneous polarity dynamics gives rise to stable polar order for $\nu\shr\tau<1$, where $p_0=\sqrt{1-\nu\shr\tau}$.

For simplicity, we focus here on fixed-magnitude polarity ($\tau\rightarrow 0$). For fixed box shear rate $\shr$ the growth rate of a solution of the linearized dynamics with co-deforming wave vector $\vec{\bar k}$ is (appendix~\ref{app: polarity only, fixed}): 
\begin{equation}
  \begin{aligned}
    &\frac{\partial_t\delta p_y(\vec{\bar k}, t)}{\delta p_y(\vec{\bar k}, t)} = \omega(\phi) \qquad \text{with}\\
    &\omega(\phi) := -\frac{1}{2}\cos{(2\phi)}\Big[\nu \cos{(2\phi)}-1\Big] + 2\nu\shr.   
  \end{aligned}
  \label{eq:growth rate, polarity only, deforming}
\end{equation}
Here, $\phi=\phi(\vec{\bar k}, t)$ is the angle of the wave vector measured in the lab frame, $\vec{k}$, which varies with time $t$ according to Eqs.~\eqref{eq:kx_kbarx} and \eqref{eq:ky_kbary} (\autoref{fig:polar-only}g,h). 
Only bend and splay modes, $\bar\phi\in\lbrace 0,\pi/2,\pi,3\pi/2\rbrace$, have angles $\phi$ that are independent of time: $\phi=\bar\phi$ (\autoref{fig:polar-only}h).

The linear stability phase diagram according to Eq.~\eqref{eq:growth rate, polarity only, deforming} is shown in \autoref{fig:polar-only}j depending on box shear rate $\shr$ and shear alignment $\nu$.
Red regions indicate unstable systems, such as systems with fixed size, $\shr=0$, as we have seen in the previous section, and freely deforming systems, $\shr = 1/4$ (green dashed line). 
%  This can be seen by inserting the box shear rate into Eq.~\eqref{eq:growth rate, polarity only, deforming}, which leads for the bend mode, $\bar\phi=\phi=0$, to a growth rate of $\partial_t\delta p_y(\vec{\bar k}, t)/\delta p_y=1/2$.
Dark gray regions indicate stable systems, which occur for deformation rates $\shr$ larger than the free deformation or for negative deformation rates $\shr<0$. The latter corresponds to the case where the boundary conditions force the system to deform perpendicular to the free deformation. 

In the yellow parameter regions in \autoref{fig:polar-only}j, the system is stable except for only transiently growing modes.
In these regions, both bend and splay modes are decaying, while a co-deforming perturbation mode with given $\vec{\bar k}$ that is neither bend nor splay mode can \emph{transiently} experience a positive growth rate. This happens as the angle of the corresponding wave vector measured in the lab frame, $\phi(\vec{\bar k}, t)$, which varies as the box is sheared, passes through a regime of angles with a positive growth rate $\omega(\phi)$ (\autoref{fig:polar-only}b,g-i).
For positive (negative) $\shr$, all angles $\phi(\vec{\bar k}, t)$ ultimately approach splay (bend) modes for $t\rightarrow \infty$, which have negative growth rates. Hence, the mode grows only transiently.

The amplitude of a transiently growing co-deforming mode will only increase by some maximal amplification factor until it decreases again (\autoref{fig:polar-only}i). This factor attains the same maximal value $S$ for all co-deforming modes $\vec{\bar{k}}$ that pass both positive zeros in the $\omega(\phi)$ curve (\autoref{fig:polar-only}b):
\begin{equation}
    S = \exp{\left[\int_{t_1}^{t_2}{\omega\Big(\phi(\bar{\phi},t)\Big)\,\d t}\right]}, \label{eq: amp factor}
\end{equation}
where $t_1$ and $t_2$ are the times when $\omega(\phi(\bar{\phi},t))$ passes zero (marked by `+' signs in \autoref{fig:polar-only}i).
% Meanwhile, co-deforming modes that start at an angle $\phi$ between both zeros will have a smaller amplification factor.
In \autoref{fig:polar-only}k, we plot the maximal amplification factor depending on $\nu$ and $\shr$.
This amplification factor is affected by two parameters, the area of the positive region in $\omega(\phi)$ (\autoref{fig:polar-only}b) and the speed by which it is traversed, which is set by the box shear rate $\shr$ (\autoref{fig:polar-only}g,h). The latter is the reason that we see a diverging amplification factor as $\shr\rightarrow 0$ (\autoref{fig:polar-only}k).
% For $\shr=0$, the area of the positive region in $\omega(\phi)$ increases if $\nu$ decreases. This is the reason why close to $\shr=0$ the amplification factor increases with decreasing $\nu$.
% Finally, for increasing $\nu\shr$, the area of the peak in $\omega(\phi)$ increases, leading to a larger amplification factor also in this regime (\autoref{fig:polar-only}i).
% However, in most of the yellow region in \autoref{fig:polar-only}j the amplification factor $S$ is not larger than 10 \mi{(SMH: false, sorry)}.

\subsubsection{Zero preferred polarity magnitude (ZPM)}
For ZPM, there is no stationary state with finite polar order in the regime where $\nu\shr\tau\geq-1$ (white region in \autoref{fig:polar-only}m).
Stabilizing polarity in these regions would require higher-order terms in the polarity free energy \cite{Giordano2021}, which we neglect for brevity here.
We focus here on the parts of the parameter space where a stationary state finite polarity magnitude $p_0$ exists, where $p_0=\sqrt{-1-\nu\shr\tau}$.
% In this regime, the stationary state of vanishing polarity (i.e.\ $\vec{p}_0=0$ and no flows) is unstable.
% In other words, a stable finite-polarity state of the homogeneous polarity dynamics only exists for deforming systems $\shr\neq 0$ with $\nu\shr<0$ (white region in \autoref{fig:polar-only}h has no stable polarity state).
% While a similar situation has been discussed in the literature before (SMH cite), we focus here on the case with a fixed box shear rate $\shr$.

We find that the linear stability diagram for systems with ZPM is quite different from systems with a fixed polarity norm (compare \autoref{fig:polar-only}j,m).
For instance, large parts of the regime with $\nu<0$ are unstable for ZPM, which is different from the case with a fixed polarity norm, where the system is stable in this regime.
This difference comes mostly from the fact that the polarity norm $p_0$ can become much larger than one for ZPM (appendix~\ref{app: polarity only, zero}). As a consequence, the perturbation growth rate is dominated by the flow created by the active stress, which scales as $\sim p_0^2$ and destabilizes polarity in this regime (appendix~\ref{app: polarity only, zero}).
% However, for smaller $\vert\nu\shr\vert$, we still find both a stable regime and a regime with transiently growing modes (\autoref{fig:polar-only}l).
Conversely, it can be shown that for $\nu\geq1$ the system is always stable (appendix~\ref{app: polarity only, zero}). 
% The system for $\nu<1$ if $\vert\shr\vert$ and $\tau$, and thus $p_0$, are large enough (\autoref{fig:polar-only}m).
% and SMH appendix figure other values of tau).
Taken together, in the regime of zero preferred polarity magnitude, the linear stability phase diagram is significantly affected by the fact that the polarity magnitude $p_0$ is influenced by the box shear rate through shear alignment.

\subsection{Comparison to system with scalar field only}
In polar-only systems (this section, \ref{sec:polar field only}), the system's behavior is symmetric with respect to whether the active stress is extensile or contractile.
%  (through the mapping  $(\alpha,\nu,\vec{p})\mapsto(-\alpha,-\nu,\bm{R}(\pi/2)\cdot\vec{p})$)
However, this was not true in scalar-only systems (\autoref{sec:scalar field only}) \cite{Tiribocchi2015a,Kirkpatrick2019a}.
% , where the system was unstable in the contractile case, while it was marginally stable in the extensile case.
Where does this difference come from?

To address this question, we map the scalar field gradient to an effective polar field $\vec{q}$ with $q_i =  \partial_i c$. This transforms the scalar field dynamics, Eq.~\eqref{eq:dimless c} into:
\begin{equation}
 \frac{\D q_i}{\D t} = -\Tilde v_{i j} q_j,
\end{equation} 
% where $\D/\D t$ denotes the co-rotational time derivative as introduced in \autoref{sec:model}.  Moreover, 
and the active stress becomes $\tilde\sigma_{ij}^a=\alpha(q_iq_j-q^2\delta_{ij}/2)$.
Thus, for an incompressible system, the scalar field dynamics corresponds to the dynamics of a polar field in the limit of no magnitude control, $\tau\rightarrow\infty$, and with a shear alignment coefficient of $+1$. 
Indeed, using our fixed-system-size results for the polar system in this limit (\autoref{fig: app fig 1}), we can retrieve our stability results for both extensile and contractile cases of the scalar system.
Hence, it is the effective positive shear alignment coefficient of $+1$ that breaks the extensile/contractile symmetry in the scalar-only system.

\section{Scalar and polar field}
\label{sec:scalar and polar field}
Here we examine the general case where the interactions between scalar, polar, and flow fields play a role.
We discuss the stability of the homogeneously deforming state given by
\begin{align}
  c_0 &= \frac{x}{l_x(t)}, \label{eq:sp c0}\\
  \vec{p}_0 &= p_0\hat{x}, \label{eq:sp p0}\\
  \v_{0,i} &= \tilde{v}_{ij}^0r_j. \label{eq:sp v0}
\end{align}
In the full system, the polarity magnitude $p_0$ generally depends on time, because it is coupled to the scalar field gradient, which is constantly becoming flatter. Thus, to simplify our discussion in this section, we choose a time-dependent $\beta$ that compensates for the flattening gradient:
\begin{equation}
  \beta(t) = \beta_0l_x(t).
\end{equation}
In this case, the state defined by Eqs.~\eqref{eq:sp c0}--\eqref{eq:sp v0} is stationary in co-deforming coordinates, where $p_0$ is implicitly set by:
\begin{equation}
  g(p_0) = \frac{\beta_0\tau}{p_0} - \nu\shr\tau. \label{eq:p0}
\end{equation}
% For both fixed and deforming systems, Eq.~\eqref{eq:p0} has a single positive solution $p_0$ for which the polarity dynamics in homogeneous shear flow is always stable.
% For fixed system size, $\shr=0$, $p_0$ depends only on $\beta_0\tau$ (\autoref{fig:scalar and polar, fixed system}b).

We prove in appendix~\ref{app: contractile coupling} that gradient-contractile systems, i.e.\ systems with $\beta>0$ and $\alpha>0$, are always unstable (\autoref{fig:scalar and polar, fixed system}a left).
% , because a $\phi=\pi/2$ mode always grows.
In the present section, we thus focus our discussion on the gradient-extensile case, $\beta>0$ and $\alpha<0$ (\autoref{fig:scalar and polar, fixed system}a right), where we show that the scalar gradient can indeed stabilize the Simha-Ramaswamy instability under certain conditions.
Such a stabilization depends in particular on how polarity magnitude is controlled; it is more effective for ZPM than for FPM.

% Moreover, we are not aware of any biological system that would correspond to the contractile case in this setting.

% Note that here the homogeneously deforming state is generally not stationary. Indeed, for a deforming system, the box shear $l_x$ changes over time, and thus also $p_0$ will change over time for $\tau>0$.
% To study the stability of the homogeneously deforming state, we will thus examine the local Lyapunov exponents of the system, a direct generalization of linear stability analysis of a stationary state.

\begin{figure*}
     \centering
     \includegraphics[width = 12.6 cm]{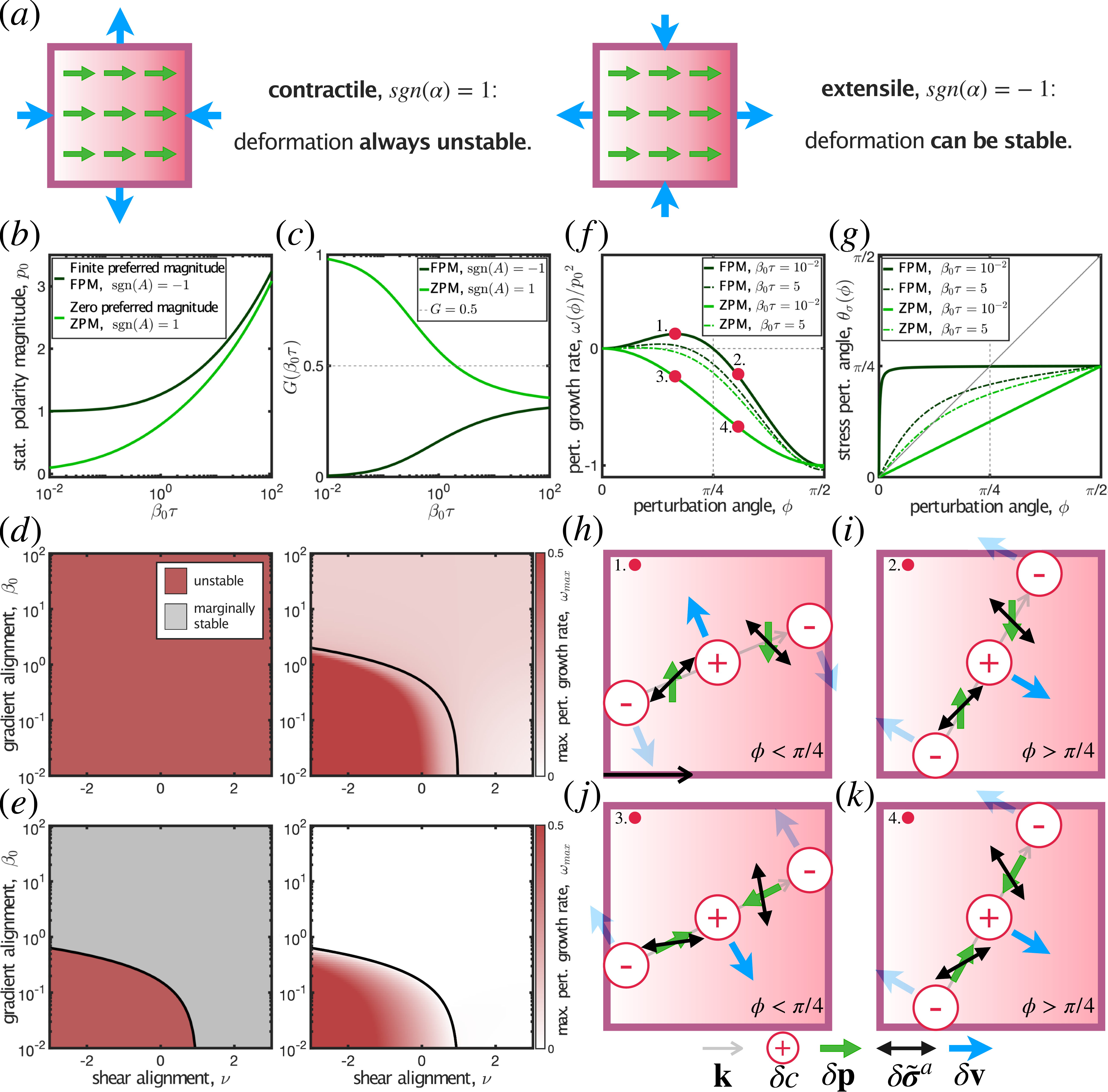}
     \caption{
     Linear stability of the non-deforming full system. A coupling to a scalar field gradient can suppress the instability of an active polar system. This stabilization is more efficient for ZPM than for FPM.
     $(a)$ Result summary:  For contractile active stresses, the system is always unstable (left). However, for extensile active stresses, the presence of the scalar field can stabilize the system (right).
     $(b)$ Stationary polarity magnitude $p_0$ vs. $\beta_0\tau$ shown for finite (dark green) and zero (light green) preferred polarity magnitude. 
     $(c)$ Stiffness ratio $G$ vs.\ $\beta_0\tau$ shown for both types of polarity. 
%      The system is unstable whenever $G<0.5$, even in the limit of strong coupling to the scalar field gradient, $\beta_0 \gg 1$. 
     $(d,e)$ Maximal perturbation growth rate $\omega_{max}$ shown depending on coupling to the scalar field gradient, $\beta_0$, and shear alignment, $\nu$, for $(d)$ fixed polarity magnitude ($\tau\rightarrow0$) and $(e)$ zero preferred polarity magnitude with $\beta_0\tau=1$. The system with fixed polarity magnitude is always unstable (red region), even for strong coupling to the scalar field gradient, whereas the system with zero preferred polarity magnitude can be marginally stabilized (light gray regions) by a strong enough coupling to the scalar field gradient $\beta_0$.
     The black solid curves indicate the places where the bend mode growth rates cross zero.
     $(f)$ Perturbation growth rate $\omega/p_0^2$ over wave vector angle $\phi$ in the limit  of strong coupling to the scalar field, $\beta_0\gg p_0^3$, shown for finite (dark green) and zero (light green) preferred polarity magnitude. Solid curves: $\beta_0\tau=10^{-2}$, dashed curves: $\beta_0\tau = 5$.
     $(g)$ Stress perturbation angle $\theta_{\sigma}$ over wave vector angle $\phi$, shown for the same cases discussed in panel e.
     $(h)$-$(k)$ Sketches illustrating the linear stability in the limit of strong coupling to the scalar field, $\beta_0\gg p_0^3$, for finite $(h)$,$(i)$  and zero $(j)$,$(k)$ preferred polarity magnitude. Perturbation angles are $0<\phi<\pi/4$ in $(h)$,$(j)$ and $\pi/4<\phi<\pi/2$ in $(i)$,$(k)$ .}
     \label{fig:scalar and polar, fixed system}
\end{figure*}

\begin{figure*}
     \centering
     \includegraphics[width = 12.6 cm]{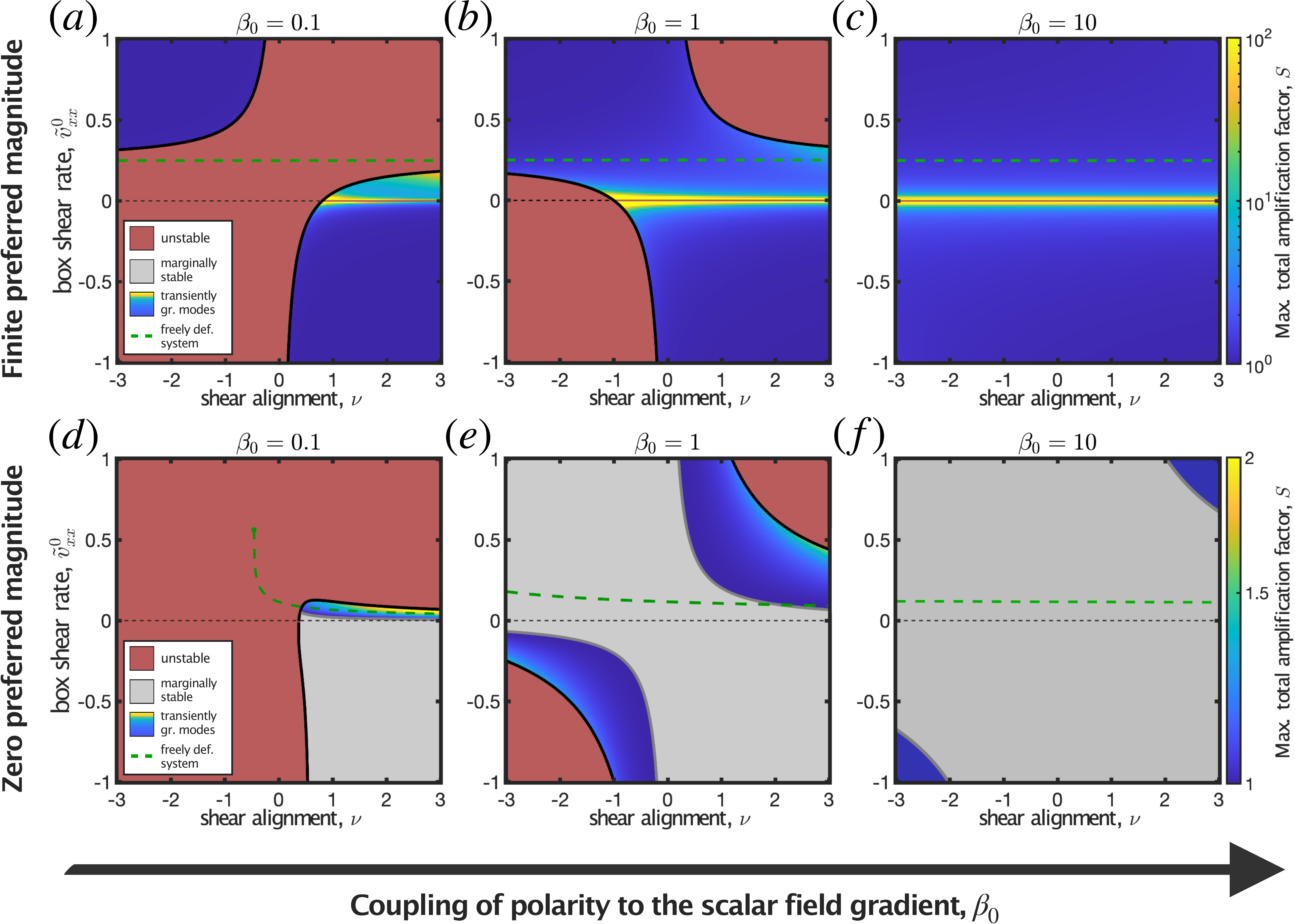}
     \caption{
     Linear stability of the deforming full system.
%      Effect of the scalar field gradient on the stability of the active polar material under deformation. 
     Stability phase diagrams depending on shear alignment $\nu$ and box shear rate $\shr$. $(a)$-$(c)$ Fixed polarity magnitude ($\tau\rightarrow0$) for $\beta_0 = (0.1,1,10)$. As the coupling $\beta_0$ to the scalar field gradient increases, the extent of unstable regions (red) shrink.
     $(d)-(f)$ zero preferred polarity magnitude ($\beta_0\tau = 1$). As $\beta_0$ increases, the extent of the marginally stable region (light gray) increases.
     The dashed green curves indicate the deformation rate of the freely deforming system. In panel d, this line is undefined for some negative $\nu$, because polarity for the freely deforming system would diverge there.
     Phase boundaries displayed as solid black curves are derived in appendix~\ref{app: analytical lines}.
     Note the different scaling of the color map between panels a-c vs.\ d-f.
     }
     \label{fig:scalar and polar, deforming}
\end{figure*}

\subsection{\texorpdfstring{Fixed system size, $\shr=0$}{Fixed system size}}
\subsubsection{Finite preferred polarity magnitude (FPM)}
\label{sec:scalar and polar, fixed, finite}
We first discuss the limit of a fixed polarity norm, $\tau\rightarrow0$, where $p_0=1$.
In this case, the system behavior depends only on two dimensionless parameters, $\nu$ and $\beta_0$.
We find that in this case the system is always unstable (\autoref{fig:scalar and polar, fixed system}d left), which we explain in the following. 

We discuss two limits, weak and strong coupling of polarity to the scalar field gradient, $\beta_0$.
Weak coupling, $\beta_0\ll p_0^3$ (appendix~\ref{app:scalar and polar field, fixed system size}), corresponds to the polarity-only case, which we discussed in \autoref{sec:polar field only}.
In this case, polarity was always unstable for fixed system dimensions (compare \autoref{fig:scalar and polar, fixed system}d right to \autoref{fig:polar-only}d).

% The maximal perturbation growth rate is largest for $\nu<0$ and small $\beta_0$ (\autoref{fig:scalar and polar, fixed system}d right), which corresponds to the growth of a bend perturbation. The region where the bend perturbation is unstable is indicated by the black curve in \autoref{fig:scalar and polar, fixed system}d right. The limit where the coupling to the scalar field gradient is weak compared to co-rotational and shear alignment effects, which means $\beta_0\ll p_0^3$ (appendix~\ref{app:scalar and polar field, fixed system size}), corresponds to the polarity-only case, where the bend instability grows fastest for $\nu<0$ (compare \autoref{fig:polar-only}d).

In the limit of strong coupling to the scalar field gradient, $\beta_0\gg p_0^3$, one might expect the same result as for the scalar-only system, \autoref{sec:scalar field only}, which was marginally stable in the gradient-extensile case.
However, we find that this is not the case, and the full system is unstable instead (\autoref{fig:scalar and polar, fixed system}d right).
To intuitively understand why, consider a given perturbation $\delta c$ of the concentration field along any wave vector with angle $0<\phi<\pi/4$ (red symbols in \autoref{fig:scalar and polar, fixed system}h). The polarity perturbation $\vec{\delta p}$ (green arrows) will adjust to $\delta c$ adiabatically quickly for strong coupling $\beta_0$.
However, because of the fixed polarity magnitude, $\vec{\delta p}$ will point in $\pm\hat{y}$ direction. As a consequence, the perturbation in the active stress tensor $\delta\tilde\sigma_{ij}^a$ is oriented along angles of $\pm \pi/4$ (black arrows, using Eq.~\eqref{eq:active stress-polarity perturbation}). The resulting flow (blue arrows) has a component that points in $-\hat{x}$ direction in regions where $\delta c$ is positive. Thus, due to convection and the gradient in $c_0(x)$, the amplitude of $\delta c$ increases.
Hence, the system is unstable for fixed polarity magnitude.

These ideas generalize to the case of a finite polarity relaxation time $\tau$.
For given scalar field perturbation $\delta c$, and strong coupling to the scalar field, $\beta_0\gg p_0^3$, the polarity perturbation $\vec{\delta p}$ relaxes adiabatically to (appendix~\ref{app:scalar and polar field, fixed system size}):
\begin{align}
  \delta p_x &= iGp_0k_x\delta c \label{eq:large-beta_delta px}\\
  \delta p_y &= ip_0k_y\delta c \label{eq:large-beta_delta py}
\end{align}
with
\begin{equation}
  G = \frac{g(p_0)}{g(p_0) + p_0g'(p_0)}.\label{eq:G}
\end{equation}
Thus, the polarity perturbation $\vec{\delta p}$ does not locally align parallel to the gradient of $\delta c$. 
Instead, the angle $\theta_p$ of the polarity perturbation is given by (precise definition of $\theta_p$ in appendix~\ref{app:scalar and polar field, fixed system size}):
\begin{equation}
  \tan{\theta_p} = \frac{1}{G}\tan{\phi}.\label{eq:theta_p}
\end{equation}
% More precisely, $\theta_p$ is defined as the angle of $\vec{\delta p}/(ik\delta c)$ (we divide by $ik\delta c$ to ensure unambigous phase).
The prefactor $G$ arises in Eqs.~\eqref{eq:large-beta_delta px}, \eqref{eq:large-beta_delta py}, because for a finite preferred polarity norm, the polarity free energy $F(p)$ panelizes perturbations $\delta p_x$ stronger than perturbations $\delta p_y$. 
In other words, $G$ is the ratio between the effective stiffnesses associated with changes in polarity away from the stationary state along $y$ and $x$ axes.
Using Eq.~\eqref{eq:p0} with $\shr=0$, the value of $G$ depends on the product $\beta_0\tau$ only (\autoref{fig:scalar and polar, fixed system}c);
for finite preferred polarity magnitude, $G$ increases from $G=0$ at $\beta_0\tau\rightarrow 0$ to maximally $G\rightarrow1/3$ at $\beta_0\tau\rightarrow\infty$.

The stiffness ratio $G$ controls the stability of the system.  To see this, we discuss the flow created by the polarity perturbation $\vec{\delta p}$, whose $x$ component is given by (appendix~\ref{app:scalar and polar field, fixed system size}):
\begin{equation}
  \delta\v_x = - p_0^2\hat{G}\,\sin{\phi}\,\sin{\big(2[\theta_\sigma-\phi]\big)}\,\delta c.\label{eq:delta vx adiabatic}
\end{equation}
Here, $\hat{G}>0$, and $\theta_\sigma$ is the angle of the active stress perturbation nematic $\delta\tilde\sigma_{ij}^a$ (precise definitions in appendix~\ref{app:scalar and polar field, fixed system size}). 
Thus, the relative direction between active stress perturbation angle $\theta_\sigma$ and wave vector angle $\phi$ determines the direction of the flow in $x$ direction.
Because the flow in Eq.~\eqref{eq:delta vx adiabatic} advects the scalar field, the system is unstable whenever $\delta\v_x/\delta c<0$. Thus,  Eq.~\eqref{eq:delta vx adiabatic} implies that the system is unstable whenever $\phi<\theta_\sigma$, where $\theta_\sigma$ can be obtained using Eq.~\eqref{eq:active stress-polarity perturbation} as: $\theta_\sigma=\theta_p/2$.

For example for fixed polarity magnitude, i.e.\ $\beta_0\tau\rightarrow 0$, we have $G=0$ (\autoref{fig:scalar and polar, fixed system}c), which implies with Eq.~\eqref{eq:theta_p} that $\theta_p=\pi/2$. Further, $\theta_\sigma=\theta_p/2=\pi/4$, and thus the modes with $\phi<\theta_\sigma=\pi/4$ are unstable (\autoref{fig:scalar and polar, fixed system}f,g).

Also for finite $\beta_0\tau$ there will always be angles $\phi$ for which $\phi<\theta_\sigma$. This is because for $\phi\ll1$, Eq.~\eqref{eq:theta_p} implies that $\theta_\sigma=\theta_p/2\simeq\phi/(2G)$. Since we have $G<1/3$  (\autoref{fig:scalar and polar, fixed system}c, appendix~\ref{app:scalar and polar field, fixed system size}), it directly follows that $\phi<\phi/(2G)=\theta_\sigma$ for small $\phi$ (see dashed dark curve in \autoref{fig:scalar and polar, fixed system}g), and thus the system is unstable. 

Taken together, for finite preferred polarity magnitude (FPM), the system is always unstable, even in the limit of strong coupling to the scalar field gradient.  This is ultimately because $G<1/3$, i.e.\ because polarity perturbation is less than $1/3$ as stiff along the $y$ axis than along the $x$ axis.

\subsubsection{Zero preferred polarity magnitude (ZPM)}
For ZPM and weak coupling to the scalar field, $\beta_0\ll p_0^3$, there is no polar-only case that we can directly compare to, because the stationary state has zero polarity in this case.
However, we find that the system is unstable for negative $\nu$ (\autoref{fig:scalar and polar, fixed system}e), which is due to an unstable bend mode (indicated by the solid black curve).
% However, as $\beta_0$ increases, there can be a large part of the parameter space where the system is marginally stable (light gray region in \autoref{fig:scalar and polar, fixed system}e), like the system without polar field in the gradient-extensile case (\autoref{sec:scalar field only}).

% To discuss how the system can be marginally stable, we focus again on the limit of $\beta_0\gg1$. For small $\beta_0\tau$, the factor $G$ as defined in Eq.~\eqref{eq:G} is close to one (\autoref{fig:scalar and polar, fixed system}b). As a consequence, the angle $\theta_p$ of the polarity perturbation follows more closely the gradient angle of the scalar field perturbation, $\phi$ (\autoref{fig:scalar and polar, fixed system}giii,iv). In this case, the angle of the stress tensor $\theta_\sigma=\theta_p/2$ is smaller than $\phi$ (see light solid curve in \autoref{fig:scalar and polar, fixed system}f), and as a consequence, using again Eq.~SMH, the flow direction $\delta\v_c\delta c$ has a component in positive $x$ direction, decreasing $\delta c$. The perturbation dynamics is jointly shown in \autoref{fig:scalar and polar, fixed system}e and illustrated in \autoref{fig:scalar and polar, fixed system}g3,g4.
% As a consequence, for small $\beta_0\tau$ the perturbation growth rate is always non-positive, and it is zero only for $\phi=0$.  Indeed, the limit $\beta_0\tau\rightarrow 0$ corresponds to the case where we only have the scalar field (see appendix~\ref{app:scalar and polar field, fixed system size}, compare Figs. SMH).

In the limit of strong coupling to the scalar field gradient, $\beta_0\gg p_0^3$, stability is again controlled by the stiffness ratio $G$.
However, because for ZPM the stiffness ratio $G$ is more isotropic, attaining values closer to one (\autoref{fig:scalar and polar, fixed system}c), the system can become marginally stable (\autoref{fig:scalar and polar, fixed system}j,k).
% In the limit of strong coupling to the scalar field gradient, $\beta_0\gg1$, the system is unstable if $G<1/2$ and stable otherwise. 
This can be seen following the same line of argument as for finite preferred magnitude. 
% In particular, for $\phi\ll 1$ Eq.~\eqref{eq:theta_p} implies again that $\theta_\sigma=\theta_p/2\simeq\phi/(2G)$, and for $G<1/2$ it follows that $\phi<\phi/(2G)=\theta_\sigma$ (\autoref{fig:scalar and polar, fixed system}g), and the system is unstable.
% Conversely, 
In particular, it can be shown that the system is marginally stable whenever $G\geq1/2$, i.e.\ whenever polarity perturbation is at least $1/2$ as stiff along the $y$ axis than along the $x$ axis (see appendix~\ref{app:scalar and polar field, fixed system size}).
This is the case whenever $\beta_0\tau\leq2$ (\autoref{fig:scalar and polar, fixed system}c,f), i.e.\ when polarity magnitude relaxation is fast enough as compared to the effect of the scalar field gradient.

% Finally, the system is not strictly stable, because at $\phi=0$, where the scalar field perturbation almost completely uncouples from the polarity perturbation, the scalar field perturbation $\delta c$ does neither grow nor shrink in the absence of diffusion (SMH Eq. scalar field dynamics, SMH appendix).

\begin{figure*}
     \centering
     \includegraphics[width = 12.6 cm]{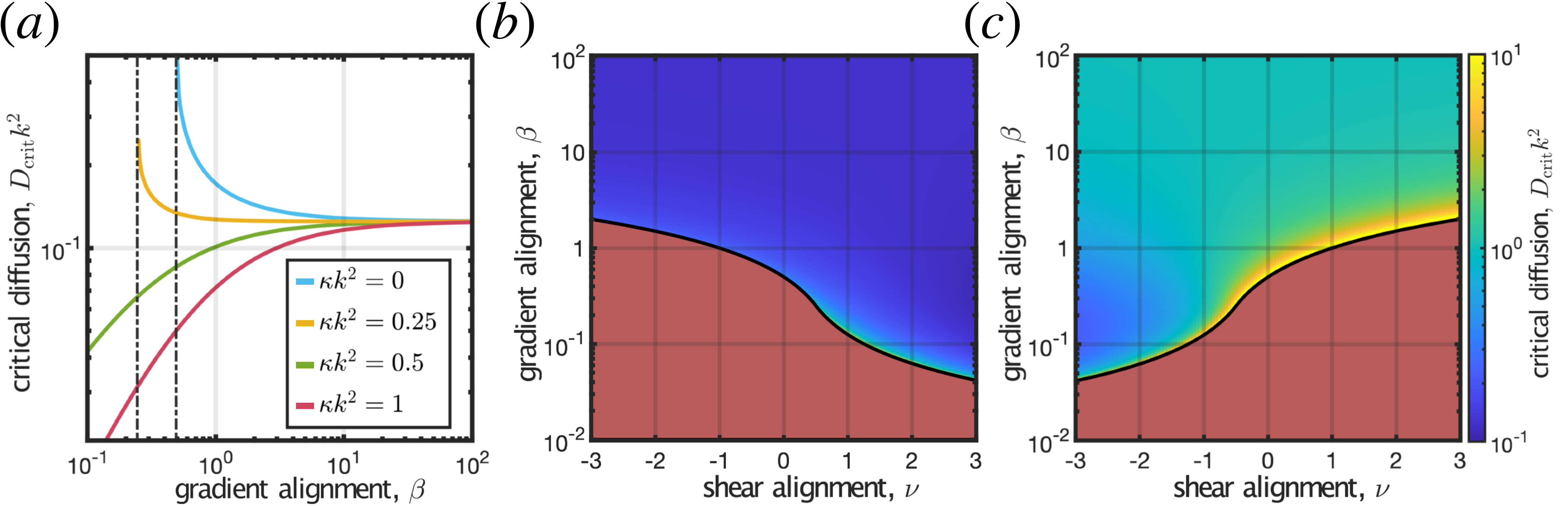}
     \caption{Stabilization of the system by diffusion and local polarity alignment, for fixed system size with fixed polarity norm. 
     (a) Minimal diffusion $D_\mathrm{crit}k^2$ required to stabilize a gradient-extensile system depending on $\beta$ and polarity alignment $\kappa k^2$.
     (b,c) Minimal diffusion $D_\mathrm{crit}k^2$ required to stabilize the system depending on $\beta$ and shear alignment $\nu$ for (b) gradient-extensile and (c) gradient-contractile systems.  In the white regions, the system cannot be stabilized by diffusion. Polarity alignment is set to $\kappa=0$.
     The color bar of panel c applies to panel b as well.
     }
     \label{fig:finite size effects}
\end{figure*}

\subsection{\texorpdfstring{Deforming system, $\shr\neq0$}{Deforming system}}
\label{sec:scalar and polar, deforming}
\subsubsection{Finite preferred polarity magnitude (FPM)}
Here we focus on the case of fixed polarity norm, $\tau=0$.
In this case, the polar-only system had two stable regions for $\nu\shr<0$ (\autoref{fig:polar-only}j).
However, when adding a weak coupling to the scalar field, $\beta_0\ll p_0^3$, these previously stable regions now turn into regions with transiently growing modes (\autoref{fig:scalar and polar, deforming}a).
This is because when including the scalar field, an additional $\omega(\phi)$ branch appears, and thus more modes that could potentially grow.
In particular, while perturbations in the polar field $\delta p_y$ relax to zero relatively quickly, any perturbation in the scalar field $\delta c$ induces a small perturbation in the polarity of $\delta p_y=-ik_y\beta_0 \delta c/\omega$. 
This polarity creates flows that then advect the scalar field. As a consequence, a scenario qualitatively similar to the one discussed in \autoref{fig:scalar and polar, fixed system}h arises. 
Here, this gives rise to growing modes with $0<\phi<\pi/4$ that are only transiently growing, because the system is deforming.
Their amplification factor is small, $S \approx 1$, because the magnitude of the growth rate is proportional to $\beta_0$ in this case.

% :  the stress tensor perturbation $\delta\tilde\sigma^a_{ij}/ik\delta c$ is has an angle of $\pi/4$, which for $0<\phi<\pi/4$ induces a flow $\delta\v_x/\delta c$ in negative $x$ direction, amplifying the scalar field.  

When increasing the coupling $\beta_0$, the region of parameter space where the system is stable with transiently growing modes is expanding (see \autoref{fig:scalar and polar, deforming}b,c).
In the limit $\beta_0\gg p_0^3$, modes with $0<\phi<\pi/4$ are again unstable for the same reason as discussed in \autoref{sec:scalar and polar, fixed, finite} (\autoref{fig:scalar and polar, fixed system}g).  In this limit, we have transiently growing modes, whose amplification factor $S$ is independent of shear alignment $\nu$ and scales as $S\sim1/\vert\shr\vert$.
% For instance, when increasing the coupling to the scalar field gradient $\beta_0$, a freely deforming system, which is unstable for small $\beta_0$, becomes stable with transiently growing modes with $S\approx1.4$ for large $\beta_0$ (green dashed lines in \autoref{fig:scalar and polar, deforming}a-c).

% Note that for negative box shear rate, $\shr<0$, the co-deforming modes $\vec{\bar{k}}$ that what we call ``transiently'' growing will actually grow forever. This is because their angle in lab coordinates $\phi(\bar{\phi},t)\rightarrow 0$ as time $t\rightarrow \infty$, while the growth rate of the mode at $\phi=0$ is zero, $\omega(\phi=0)=0$.  However, the total amplification factor $S$ in these cases is still finite (SMH FIg, see appendix~\ref{app: finite amp. factor})

\subsubsection{Zero preferred polarity magnitude (ZPM)}
% For zero preferred polarity magnitude, the presence of a coupling to a scalar field gradient first leads to a finite polarity magnitude in the whole $(\nu,\shr)$ parameter space (\autoref{fig:scalar and polar, deforming}d-f), in contrast to the polarity-only case (\autoref{fig:polar-only}h).
For ZPM and weak coupling to the scalar field gradient, $\beta_0\ll p_0^3$, stability is somewhat similar to the polarity-only case (compare \autoref{fig:polar-only}m, \autoref{fig:scalar and polar, deforming}d).
The main difference is that most of the regions of parameter space without stationary state in the polarity-only system become unstable when adding a weak coupling to a scalar field gradient.
Increasing the coupling to the scalar field $\beta_0$ generally leads to an expansion of the parameter regime with marginal stability (see \autoref{fig:scalar and polar, deforming}e,f).
% For instance, the freely deforming system is unstable or has transiently growing modes for small $\beta_0$, while it becomes marginally stable for large $\beta_0$ (green dashed curves in \autoref{fig:scalar and polar, deforming}d-f).

\section{System size effects}
\label{sec:system size effects}
Until now we have discussed the limit of infinite system size, where neither diffusion $D$ nor polarity alignment $\kappa=K/\gamma$ can suppress large-wavelength perturbations with $k\rightarrow0$.
However, a finite system size limits the wavelength from above, which can allow diffusion and polarity alignment to stabilize the system. 
Here we discuss under which conditions they can do so, focusing on the case of a fixed system size and a fixed polarity magnitude. In this case, the system was always unstable in the infinite-system-size limit (\autoref{sec:scalar and polar, fixed, finite}).

For the limits of scalar-only and polar-only dynamics, the stability criteria are already known.
For the scalar-only limit, an otherwise unstable gradient-contractile system is stabilized whenever \cite{Kirkpatrick2019a}
\begin{equation}
Dk^2 > 1,
\end{equation}
where we use dimensionless units (\autoref{sec:dimless}), i.e.\ rates are in units of $\vert\alpha\vert/\eta$.
For the polar-only limit, for example in the extensile case, local polarity alignment can stabilize the system only when \cite{Voituriez2005}
\begin{equation}
 \kappa k^2 > \begin{cases}
    \frac{1}{2}(1-\nu) & \text{for $\nu\leq1/2$ and} \\ 
    1/(8\nu) & \text{for $\nu>1/2$.}
    \end{cases}
\end{equation}

For the full system with interacting scalar and polar fields, we derive the minimal diffusion term required to stabilize the system in appendix~\ref{app:system-size effects}.  For instance, for the gradient-extensile case we find that diffusion can stabilize the system only if
\begin{equation}
  \beta+\kappa k^2 >
    \begin{cases}
      \frac{1}{2}(1-\nu) & \text{for $\nu\leq1/2$ and} \\
      1/(8\nu) & \text{for $\nu>1/2$.}
    \end{cases}\label{eq:diffusion-criterion-gradient-extensile}
\end{equation}
This means that for diffusion to stabilize the system, the gradient alignment $\beta$ needs to be strong enough to dominate the polar field dynamics.
If Eq.~\eqref{eq:diffusion-criterion-gradient-extensile} is fulfilled, a finite value of the diffusion constant can stabilize the system: $Dk^2>D_\mathrm{crit}k^2>0$ (the analytical expression for $D_\mathrm{crit}k^2$ is reported in appendix~\ref{app:system-size effects}).
Thus, while those systems that fulfill Eq.~\eqref{eq:diffusion-criterion-gradient-extensile} can be stabilized by diffusion alone (i.e.\ with $\kappa=0$), a system can never be stabilized by polarity alignment alone (i.e.\ with $D=0$). This is due to the instability that arises from the coupling of scalar and polar fields (see adiabatic limit discussed in \autoref{sec:scalar and polar, fixed, finite}). We find that this is also true in the gradient-contractile case (appendix~\ref{app:system-size effects}).

In \autoref{fig:finite size effects}, we report the magnitude of $D_\mathrm{crit}k^2$ for varying $\beta$, $\nu$, and $\kappa k^2$.
For instance, for $\nu=0$ and without polarity alignment, $\kappa=0$, the critical diffusion $D_\mathrm{crit}k^2$ decreases monotonically with the alignment of polarity to the scalar field, $\beta$ (blue solid curve in \autoref{fig:finite size effects}a), which intuitively makes sense since diffusion and $\beta$ both act together to stabilize the instability of the polar-only system. However, for larger values of the polarity alignment $\kappa k^2$, we find that $D_\mathrm{crit}k^2$ can also monotonically increase with $\beta$ (red and green solid curves in \autoref{fig:finite size effects}a). In this regime, the polar-only limit is not unstable any more, and diffusion is required to compensate for the instability that arises from the coupling of scalar and polar field (\autoref{sec:scalar and polar, fixed, finite}).

Finally, we summarize the stability criteria for the gradient-extensile and gradient-contractile cases in \autoref{fig:finite size effects}b and c, respectively.  As expected, the critical diffusion term is generally larger in the gradient-contractile case than in the gradient-extensile case.

\section{Discussion}
\label{sec:discussion}
The homogeneously deforming state of polar or nematic wet active matter is subject to the well-known Simha-Ramaswamy instability \cite{Simha2002,Voituriez2005}.
This raises the question how, despite this instability, active anisotropic tissue deformation can be robust during animal morphogenesis \cite{Wolpert2015}.
Animal morphogenesis is known to be organized by large-scale protein concentration patterns (e.g.\ morphogen gradients) \cite{Wolpert2015}. Under which conditions could such patterns stabilize anisotropic tissue deformation?
To address this question, we examined whether a scalar field gradient can stabilize the homogeneously deforming state of an active polar material, even when the scalar field is advected by material flows.

Focusing on the limit of a large system size, we showed that the homogeneously deforming state is always unstable in gradient-contractile systems, i.e.\ when the active anisotropic stress is, mediated by the polarity, contractile along the direction of the gradient. However, the system can be marginally stable in the gradient-extensile case.
This is true both when the active anisotropic stress is controlled directly by the scalar field gradient (\autoref{sec:scalar field only}, \autoref{fig:scalar field only}) and when this control is mediated through the polar field (\autoref{sec:scalar and polar field}, \autoref{fig:scalar and polar, fixed system}).
Intriguingly, in the biology literature we found many instances of animal morphogenetic systems where the effective coupling between controlling morphogen gradient and tissue deformation is gradient-extensile \cite{Kong2017,Bosveld2012,Johansen2003,Ninomiya2004,Tao2019,Saxena2014,Alegot2018,Benazeraf2010}.
However, so far we could identify almost no instance of a gradient-contractile coupling.
This predominance of gradient-extensile systems in developing animals has been remarked once before in the biology literature \cite{Keller2006}, and our results provide an explanation purely based on active matter physics.
% Our results provide an explanation for this, suggesting that gradient-extensile coupling may be preferred by evolution because it would lead to more stable tissue deformation.

The only example we have found where something akin to a gradient-contractile coupling has been proposed is convergent-extension of the notochord in the ascidian \textit{Ciona intestinalis} \cite{Shi2009}. 
However, evidence is sparse, and the deforming tissue is very small consisting only of 40 cells, which would facilitate other mechanisms of stabilization (see below).

In our study, gradient-extensile systems can only become \emph{marginally} stable.  They do not become strictly stable because for perturbations parallel to the gradient direction, the scalar field decouples from flow, and scalar field perturbations along this direction are marginal, i.e.\ they do neither grow nor shrink.
This will be different in more realistic systems, where protein gradients are created, e.g., by secretion, diffusion, and degradation \cite{Turing1952,Wolpert2015,Romanova-Michaelides2022}.
% (SMH: instead of diffusive morphogen gradient, could be cell identity graded along on axis but not the other, that was maybe defined by an earlier gradient, e.g. drosophila germ band)

We further showed that the stability of gradient-extensile systems strongly depends on how the polarity \emph{magnitude} is controlled, where we compare the cases of finite and zero preferred magnitude (\autoref{sec:scalar and polar field}, \autoref{fig:scalar and polar, fixed system},\ \ref{fig:scalar and polar, deforming}).
Both kinds have examples in biological tissues. 
Finite-preferred-magnitude (FPM) polarity resembles Core/Frizzled planar cell polarity (PCP), which is believed to emerge without any cell-external cues \cite{Devenport2014}.
Meanwhile, zero-preferred-magnitude (ZPM) polarity resembles Fat PCP \cite{Brittle2012} and actin polarity in the \textit{Drosophila} germ band \cite{Zallen2004,Lavalou2021}, which are believed to show no polarization in the absence of external cues.
% So far, the influence of polarity magnitude has often been an undervalued aspect in its influence on active matter dynamics.  % overlooked, undervalued, neglected
We demonstrate that for FPM, the response of polarity perturbations to scalar field perturbations is more anisotropic than for ZPM, which affects the generated active stresses and material flows.
As a consequence, systems with ZPM are generally more stable than systems with FPM.
However, in both cases the coupling of the polar field to the scalar field gradient generally has a stabilizing effect.

% Active polar systems under pure shear deformation can still be stabilized by the shear alignment effect without scalar field.
% This can occur when external stresses deform the active material faster than the material would deform in the absence of external stresses (free deformation), which could be relevant for \textit{Drosophila} germ band extension \cite{Collinet2015}.
% Active polar systems could also be stabilized when external forces deform the active material opposite to the direction in which the material deforms without external stresses.  This could be relevant for \textit{Drosophila} wing development \cite{Etournay2015,Dye2021}.

We also demonstrated the emergence of parameter regions with transiently growing perturbation modes. 
These arise because for systems under pure shear deformations the solutions to the linearized dynamics are co-deforming modes, whose wave vectors are changing during the overall system deformation. As a consequence, the growth rate of the perturbation modes can change over time.
We identified parameter regions where some perturbation modes decay, while for other modes the amplitude only transiently grows before decaying forever.
To characterize the stability of the system in these regions, we introduce a maximal amplification factor $S$, which these modes experience while their amplitude grows. 

% We showed that the instability in the contactile case of the scalar-only system is analogous to the known Simha-Ramaswamy instability of the polar-only system. This can be seen directly by mapping gradients of the scalar field to an effective polar field (\autoref{sec:}). The asymmetry between contractile (unstable) and extensile (marginally stable) cases is then explained by an effective shear alignment coefficient of $\nu=+1$.

While we discussed here one potential mechanism to stabilize the Simha-Ramaswamy instability during animal morphogenesis, there may be other mechanisms as well.
First, in most of our article we focused on the infinite system size limit.  For finite polar-only systems, polarity alignment $K$, possibly in combination with the boundary conditions for the polarity, can also stabilize the system. There is a maximal tissue size scale $L_c\sim\sqrt{K/\gamma\tilde v}$, where $\tilde v$ is the free tissue deformation rate, beyond which polarity alignment becomes insufficient to stabilize the system \cite{Voituriez2005} (\autoref{sec:system size effects}).
Combining known orders of magnitude for active fly wing deformation $\tilde v\sim(10^{-2}\dots10^{-1})\,\hour^{-1}$ \cite{Etournay2015}, and a measured PCP alignment rate in the wing, $K/\gamma\sim(1\dots10)\,\micro\metre^{2}/\hour$ \cite{Merkel2014}, we obtain $L_c\sim10\,\micro\metre$, which is on a similar order as values measured in monolayers of different mouse cell lines of $L_c\sim40\,\micro\metre$ \cite{Duclos2018}.
However, many developing tissues have sizes of $\sim(10\dots100)\,\micro\metre$, suggesting that an effective cell polarity alignment is not necessarily sufficient to stabilize the Simha-Ramaswamy instability.
When including diffusion and polarity alignment in our calculations for the full system of interacting scalar and polar fields, we find that when scalar and polar fields interact, polarity alignment alone is never sufficient to stabilize the system, at least for fixed polarity norm, while diffusion can be (\autoref{sec:system size effects}). This is interesting given that there are biological systems where diffusion is not believed to occur, such as during \textit{Drosophila}\ germ band extension.

Second, the Simha-Ramaswamy instability occurs only in wet active matter, i.e.\ when momentum (and angular momentum \cite{Maitra2018}) is conserved.  In other words, active tissue deformation could also be stabilized, e.g.\ by friction $\zeta$ with a substrate whenever the the hydrodynamic length scale $L_h=\sqrt{\eta/\zeta}$ is sufficiently small as compared to $L_c$.
This could be relevant for instance for \textit{Drosophila} germ band extension \cite{DAngelo2019}. However, many morphogenetic tissue deformation processes, may be better described as wet active matter.  This includes for instance vertebrate limb bud elongation \cite{Sermeus2022} and morphologically similar processes \cite{Tao2019}, as well as hydra morphogenesis \cite{Maroudas-Sacks2021}.

Third, active oriented materials may be stabilized by lifting the condition of incompressibility \cite{Maitra2021}.
Indeed, developing tissues can show some degree of compressibility.
For instance a finite tissue bulk viscosity can arise from cell division and death \cite{Ranft2010}. 
However, such a bulk viscosity may become visible only on time scales above the cell division time, which is typically on the order of hours or days. This is also the approximate range of typical anisotropic tissue deformation processes during development.
Thus, a bulk viscosity due to cell division would be relevant in particular for slow tissue deformation processes.
Moreover, layered 2D tissues called epithelia may additionally exhibit limited 2D compressibility through variation of layer height or cell extrusion.
However, a finite compressibility alone is not necessarily sufficient to stabilize deforming systems. For instance, while for a polar-only system with fixed size, a finite bulk viscosity can indeed stabilize the system for sufficiently large value of $\nu$, this stabilizing effect only acts on angles other than multiples of $\pi/2$, i.e.\ it does not act on bend or splay modes.  Consequentially, for deforming systems, a finite bulk viscosity would help stabilize regions with transiently growing modes (e.g.\ yellow regions in \autoref{fig:polar-only}j), but it would not stabilize unstable regions (e.g.\ red regions in \autoref{fig:polar-only}j). For instance, a freely deforming polar-only system is unstable also if the tissue is compressible (green dashed line in \autoref{fig:polar-only}j).

% It is so far unclear in how far such processes could contribute to the stabilization of anisotropic tissue deformation.

% There is also angated, providing a nematic order parameter that interacts with tissue mechanics \cite{Etournay2015,Guirao2015,Merkel2017,Popovic2017}.

Our work prompts for different kinds of experiments on animal morphogenetic systems to test our ideas.
First, are really all active anisotropic tissue deformation processes gradient-extensile?
So far, there are many systems where the precise role of morphogens for tissue deformation is still unknown \cite{Maroudas-Sacks2021,Sermeus2022,Etournay2015,Dye2021,Williams2020}.
In many systems, it is believed that scalar field gradients control the orientation anisotropic deformation, but more experimental evidence is required.
Also, is notochord convergent-extension of \textit{Ciona intestinalis} indeed a counter example, i.e.\ a gradient-contractile system?  If so, it is possibly not the only one.  How is tissue deformation stabilized in these systems?

Second, relatively little is currently known about what kinds of polarity are used to control tissue deformation, i.e.\ whether they are ZPM or FPM polarity.
In the few systems where more is known, it appears that ZPM polarity controls tissue deformation \cite{Bosveld2012,Kong2017}, which in our analysis leads to a more robust behavior.
Is this kind of polarity indeed more often used to control tissue deformation in morphogenesis?

Third, many parameter values are still unknown, even in the best-studied biological systems.  For instance, while something like a shear alignment effect has been observed in a few systems now \cite{Aigouy2010,Merkel2014,Merkel2014b,Chien2015,Aw2016,Blanch-Mercader2017c,Duclos2018,Blanch-Mercader2021}, we are aware of only two systems where the shear alignment parameter $\nu$ has been measured, the \textit{Drosophila} wing \cite{Aigouy2010,Merkel2014,Merkel2014b} and certain cell monolayers in vitro \cite{Blanch-Mercader2017c,Duclos2018,Blanch-Mercader2021}.
Measuring parameters like this in more developmental systems will allow to quantitatively test our predictions.

While a foundational motivation of active matter physics has always been to better understand collective motion in living systems \cite{Bowick2022a}, a lot remains to be learned at the direct interface with biology.
Here, we provide an example for how active matter physics may reveal fundamental principles for animal morphogenesis.
% We hope that our results will soon be tested by new experimental data.
% testing these principles.

\begin{acknowledgments}
We thank the Centre Interdisciplinaire de Nanoscience de Marseille (CINaM) for providing office space and S. Gsell for a critical reading of the manuscript.
The project leading to this publication has received funding from France 2030, the French Government program managed by the French National Research Agency (ANR-16-CONV-0001), and from the Excellence Initiative of Aix-Marseille University - A*MIDEX.
\end{acknowledgments}

\appendix

\section{Co-deforming coordinates}
\label{app: co-deforming coordinates}
\subsection{Definition}
\label{app:co-deforming coordinates definition}
In order to more conveniently solve the linearized dynamics (appendix~\ref{app:linearized dynamics}), we introduce co-deforming coordinates $(\bm{\bar r},\bar t)=(\bar x, \bar y, \bar t)$. These coordinate map to the lab coordinates $(\bm{r}, t)=(x, y, t)$ in the following way:
\begin{align}
  r_i &= s_{ij}(\bar{t})\bar{r}_j \label{eq:co_deforming r}\\
  t &= \bar{t}, \label{eq:co_deforming t}
\end{align}
where $\bm{s}(\bar{t})$ is a time-dependent shear tensor, given by
\begin{equation}
  \bm{s}(\bar{t}) = \begin{pmatrix}
                l_x(\bar{t}) & 0 \\ 0 & l_x^{-1}(\bar{t})
              \end{pmatrix}. \label{eq:shear tensor}
\end{equation}
Thus, while at some time $t$, lab coordinates range from $0\leq x<L_x(t)$ and $0\leq y<L_y(t)$, co-deforming coordinates map these affinely to the box dimensions at time zero, with $0\leq \bar{x}<L_x(0)$ and $0\leq \bar{y}<L_y(0)$.

As a direct consequence of Eqs.~\eqref{eq:co_deforming r} and \eqref{eq:co_deforming t}, partial derivatives of some quantity $q$ transform as:
\begin{align}
  \bar\partial_jq &:= \frac{\partial q(\bm{\bar{r}}, \bar{t})}{\partial \bar{r}_j} = (\partial_iq)s_{ij} \label{eq:trafo partial derivative r}\\
  \bar\partial_tq &:= \frac{\partial q(\bm{\bar{r}}, \bar{t})}{\partial \bar{t}} = \partial_tq + (\partial_iq)\dot{s}_{ij}\bar{r}_j, \label{eq:trafo partial derivative t}
\end{align}
where $\partial_iq := \partial q(\bm{r}, t)/\partial r_i$, $\partial_tq := \partial q(\bm{r}, t)/\partial t$, and $\dot{s}_{ij} := \d s_{ij}/\d t= \d s_{ij}/\d\bar{t}$. Thus, the partial time derivative in co-deforming coordinates, $\partial_{\bar{t}}q$, i.e.\ for fixed $\bm{\bar{r}}$, includes a term related to the box shear rate as compared to the partial time derivative with respect to lab coordinates.

Moreover, with the co-deforming Fourier transformation of a quantity $q$ defined as in Eq.~\eqref{eq:codef fourier}, we have the usual derivation rule, where the Fourier transform of $\bar\partial_j q(\bm{\bar r}, \bar t)$ is $i\bar{k}_jq(\bm{\bar k}, \bar t)$.
From Eqs.~\eqref{eq:codef fourier} and \eqref{eq:co_deforming r} also follows that a given co-deforming Fourier mode with wave vector $\bm{\bar{k}}$ corresponds to a lab frame Fourier mode with wave vector $\bm{k}$ with components
\begin{equation}
  k_i = \bar{k}_js^{-1}_{ji}, \label{eq:trafo codef k}
\end{equation}
because then we have $\bm{\bar{k}}\cdot\bm{\bar{r}} = \bm{k}\cdot\bm{r}$.
% Compare also Eq.~\eqref{eq:trafo codef k} with Eqs.~\eqref{eq:kx_kbarx}, \eqref{eq:ky_kbary}, and Eq.~\eqref{eq:trafo partial derivative r}.

\subsection{Velocity}
To obtain the mapping for the velocity field, we consider a tracer particle that is perfectly advected with the flows.  The velocity of that tracer particle corresponds to a total time derivative $\v_i=\d r_i/\d t$, for which we obtain by insertion of Eq.~\eqref{eq:co_deforming r}:
\begin{equation}
  \v_i = \dot{s}_{ij}\bar{r}_j + s_{ij}\bar{\v}_j, \label{eq:co-deforming v}
\end{equation}
where $\bar\v_i=\d \bar{r}_i/\d t=\d \bar{r}_i/\d \bar{t}$ is the co-deforming velocity, with $\bar{r}(\bar{t})$ being the tracer trajectory in co-deforming coordinates.
The first term in Eq.~\eqref{eq:co-deforming v} corresponds to a motion due to the affine transformation according to box coordinates. Thus, $\bar\v_i$ can be interpreted as the non-affine component of the flow field.

To obtain a transformation formula for the convective derivative, we consider again our tracer and the presence of some spatio-temporal field $q$.  The convective derivative corresponds to the total derivative of the value of $q$ that the tracer locally sees. Thus, we expect analogous expressions for the convective derivative in both lab and co-deforming systems, $\dot{q} := \d q/\d t= \d q/\d \bar t$. Indeed, using Eqs.~\eqref{eq:trafo partial derivative r}--\eqref{eq:co-deforming v}, we obtain:
\begin{equation}
  \dot{q} = \partial_tq + \v_i(\partial_i q) = \bar\partial_tq + \bar\v_i(\bar\partial_i q).
\end{equation}

\subsection{Dynamical equations}
The dynamical equations for scalar and polar fields, Eqs.~\eqref{eq:dimless c}--\eqref{eq:dimless p}, in co-deforming coordinates are: 
\begin{align}
  \frac{\d c}{\d \bar t} &= 0 \label{eq:codef c}\\
  \frac{\d p_i}{\d \bar t} &= -\frac{g(p)}{\tau}p_i -  \nu\tilde{v}^0_{ij}p_j + \beta s^{-1}_{ij}\bar\partial_j c \nonumber \\
  &\quad - \frac{1}{2}\bigg[(\nu+1)s^{-1}_{li}s_{jk} + (\nu-1)s^{-1}_{lj}s_{ik}\bigg]\bar\partial_l\bar\v_kp_j\label{eq:codef p}
%   \\
%   0 &= \partial_i^2\v_j - \partial_j\Pi' + \sgn{\alpha}\,\partial_i(p_ip_j) \label{eq:dimless v}\\
%   \bar\partial_i\bar\v_i &= 0. \label{eq:codef incompressibility}
\end{align}
Here, we left out diffusion and polarity alignment, and we use the box shear rate $\tilde{v}^0_{ij}=(\dot{s}_{jk}s^{-1}_{ki} + \dot{s}_{ik}s^{-1}_{kj})/2$, which is for the box shear tensor defined in Eq.~\eqref{eq:shear tensor}:
\begin{equation}
 \bm{\tilde{v}^0} = \begin{pmatrix}
                      \dot{l}_x/l_x & 0 \\ 0 & -\dot{l}_x/l_x
                    \end{pmatrix}.
\end{equation}
We do not rewrite in co-deforming coordinates the incompressible Stokes' equations, Eqs.~\eqref{eq:dimless v} and \eqref{eq:dimless incompressibility}, because this will not be required in what follows.

\subsection{Linearized dynamics around the homogeneously deforming state}
\label{app:linearized dynamics}
We linearize the dynamics around the homogeneously deforming state, given by
\begin{align}
  c_0 &= \xb &
  \vec{p}_0 &= p_0\bm{\hat{x}} &
  \vec{\vb}_0 &= \vec{0}. \label{eq:homogeneously deforming state}
\end{align}
Using co-deforming coordinates will facilitate dealing with the advective terms when solving the linearized dynamics.

To fix a value for $p_0$ at some time point $t$, we use Eqs.~\eqref{eq:codef p} and \eqref{eq:homogeneously deforming state} with the stationarity condition $\d p_x/\d t=0$:
\begin{equation}
  g(p_0) = \frac{\beta\tau}{p_0l_x(t)} - \nu\shr\tau. \label{eq:stationary p0}
\end{equation}
For constant $\beta$ and a deforming box, the state $p_0$ given by this equation is only \emph{transiently} stationary, due to the time-dependent $l_x$.
However, in the adiabatic limit where polarity relaxation is much faster than box deformation, $\tau\vert\shr\vert\ll1$, the homogeneous dynamics will generally be close to the state $p_0$ given by Eq.~\eqref{eq:stationary p0}.
In the main text, we circumvent these issues by setting
\begin{equation}
  \beta(t) = \beta_0l_x(t).
\end{equation}
with constant $\beta_0$. In this case Eq.~\eqref{eq:stationary p0} always defines a stationary solution of the dynamics.

We consider the following perturbation to linear order:
\begin{align}
  c &= \xb + \delta c \\
  \vec{p} &= p_0\bm{\hat{x}} + \vec{\delta p} \\
  \vec{\vb} &= \vec{\delta\vb}.
\end{align}
Insertion into Eqs.~\eqref{eq:codef c} and \eqref{eq:codef p} yields to first order in co-deforming Fourier space:
\begin{align}
  \bar\partial_t\delta c(\bm{\bar k}, \bar t) &= -\delta \vb_x \label{eq: bpartial delta c}\\
  \bar\partial_t \delta p_x(\bm{\bar k}, \bar t) &= 
    \left(-\frac{1}{\tau}\Big[g(p_0) + g'(p_0)p_0\Big] -\nu \shr \right)\delta p_x \nonumber\\
    &\quad+i\beta l_x^{-1}\bar{k}_x\delta c
    -i\nu p_0 \bar{k}_x \delta \vb_x \label{eq: bpartial delta px}\\
  \bar\partial_t \delta p_y(\bm{\bar k}, \bar t) &= 
    \left( -\frac{1}{\tau}g(p_0) +\nu\shr \right) \delta p_y 
    + i\beta l_x\bar{k}_y\delta c \nonumber\\
    &\quad-\frac{ip_0}{2}\Big[(\nu + 1)l_x^2\bar{k}_y\delta\vb_x + (\nu - 1)l_x^{-2}\bar{k}_x\delta\vb_y\Big]
    .
    \label{eq: bpartial delta py}
\end{align}
Using the co-deforming coordinates allowed us to simplify the advective terms, which otherwise lead to spatially dependent coefficients. Here, for simplicity, we left out the diffusion and local polarity alignment terms.

To close the system of equations, Eqs.~\eqref{eq: bpartial delta c}--\eqref{eq: bpartial delta py}, we need to obtain an expression for $\bm{\delta\vb}$ by solving the incompressible Stokes' equations, Eqs.~\eqref{eq:dimless v} and \eqref{eq:dimless incompressibility}.  We insert the expression for the lab-frame velocity Eq.~\eqref{eq:co-deforming v} into these equations, and obtain after lab-frame Fourier transformation of the linearized dynamics:
\begin{equation}
  \delta\vb_m = \frac{is^{-1}_{mi}k_l}{\eta k^2} \left( \delta_{ij} - \frac{k_i k_j}{k^2} \right)\delta\tilde\sigma^a_{lj}.
  \label{eq:co-deforming velocity}
\end{equation}
Here, $k$ is the magnitude of the lab-frame wave vector $\vec{k}$. Using that to linear order the active stress $\delta\tilde\sigma^a_{lj}$ is given by Eq.~\eqref{eq:active stress-polarity perturbation}, we obtain:
\begin{align}
  \delta {\vb}_x &= \frac{il_x^{-1}}{k} \sgn{\alpha}p_0\sin{(\phi)} \delta\Phi \label{eq: codeforming delta vx} \\
  \delta {\vb}_y &= -\frac{il_x}{k}\sgn{\alpha}p_0\cos{(\phi)} \delta\Phi \label{eq: codeforming delta vy}
\end{align}
with $\delta\Phi = \sin{(2\phi)} \delta p_x - \cos{(2\phi)} \delta p_y$, and $\phi$ being the angle of the lab-frame wave vector $\bm{k}$.
Note that a transformation of Eqs.~\eqref{eq:dimless v} and \eqref{eq:dimless incompressibility} into co-deforming coordinates, followed by a linearization and a co-deforming Fourier transformation leads to the same result.
% To linear order, the active stress is given by $\delta\tilde\sigma^a_{xx}=-\delta\tilde\sigma^a_{yy}=\sgn{\alpha}p_0\delta p_x$ and $\delta\tilde\sigma^a_{xy}=\delta\tilde\sigma^a_{yx}=\sgn{\alpha}p_0\delta p_y$.

\begin{widetext}
Inserting the velocity perturbation, Eqs.~\eqref{eq: codeforming delta vx} and \eqref{eq: codeforming delta vy}, into the linearized dynamics, Eqs.~\eqref{eq: bpartial delta c}--\eqref{eq: bpartial delta py}, we obtain:
\begin{align}
\bar\partial_t \delta c &= -Dk^2\delta c
-\frac{il_x^{-1}}{k}\sgn{\alpha}p_0 \sin{\phi}\sin{2\phi} \;\delta p_x 
+ \frac{il_x^{-1}}{k}\sgn{\alpha}p_0 \sin{\phi}\cos{2\phi} \;\delta p_y 
\label{eq:linearized dynamics delta c} \\
\bar\partial_t \delta p_x &= 
i\beta k \cos{\phi} \; \delta c 
+ \left[ -\frac{1}{\tau} \Big( g(p_0) + g'(p_0)p_0 \Big) -\nu\shr + \frac{\sgn{\alpha}p_0^2}{2}\nu\sin^2{2\phi} - \kappa k^2\right] \delta p_x 
-\frac{\sgn{\alpha}p_0^2}{2}\nu \sin{2\phi}\cos{2\phi} \;\delta p_y 
\label{eq:linearized dynamics delta px} \\
\bar\partial_t \delta p_y &= 
i\beta k \sin{\phi} \;\delta c 
- \frac{\sgn{\alpha}p_0^2}{2} \sin{2\phi}\big(\nu \cos{2\phi}-1\big) \;\delta p_x \nonumber\\
&\qquad\qquad\qquad\qquad\qquad\qquad\qquad\qquad
+ \left[- \frac{1}{\tau}g(p_0) + \nu\shr + \frac{\sgn{\alpha} p_0^2}{2}\cos{2\phi}\big(\nu \cos{2\phi}-1\big) - \kappa k^2\right]\delta p_y.  \label{eq:linearized dynamics delta py}
\end{align}
The derivatives on the left-hand sides are partial derivatives for constant co-deforming wave vectors $\bm{\bar k}$. Thus, solutions to the linearized dynamics are co-deforming Fourier modes with time-dependent amplitude.  The right-hand side is written in terms of angle $\phi$ and magnitude $k$ of the lab-frame wave vector out of convenience only.
\end{widetext}

\section{Linear stability of the homogeneously deforming state}
Here, we discuss the linear stability of the homogeneously deforming state based on the linearized dynamics, Eqs.~\eqref{eq:linearized dynamics delta c}--\eqref{eq:linearized dynamics delta py}.

\subsection{\texorpdfstring{Gradient-contractile systems ($\sgn{\alpha} = 1$) are always unstable.}{Gradient-contractile systems are always unstable.}}
\label{app: contractile coupling}
Here we show that the system is always unstable for $\sgn{\alpha} = 1$ and $\beta>0$, and for vanishing diffusion and polarity alignment. This is because in this case there is a co-deforming mode with wave vector direction $\bar\phi = \pi/2$ that is permanently growing. For this specific angle, this corresponds to a lab-frame wave vector with the same angle, $\phi=\pi/2$.

To prove that there is a mode with $\phi=\pi/2$ that is permanently growing, we show that for this angle the characteristic polynomial $P(\omega)$ of the matrix describing the linearized dynamics, Eqs.~\eqref{eq:linearized dynamics delta c}--\eqref{eq:linearized dynamics delta py}, has at least one positive zero. For $\phi = \pi/2$ this polynomial is:
\begin{equation}
\begin{split}
  P(\omega) &= \sgn{\alpha}\beta l_x^{-1} \left[\beta l_x^{-1}+ \frac{1}{\tau}g'(p_0)p_0^2\right] \\
  &\quad+\sgn{\alpha}\beta l_x^{-1}p_0\omega \\
  &\quad-\omega \left[-\frac{1}{\tau} \Big(g(p_0) +  g'(p_0)p_0\Big) -\nu\shr -\omega\right] \\ 
  &\qquad\times \left[- \frac{1}{\tau} g(p_0)  + \nu\shr + \frac{\sgn{\alpha}p_0^2}{2} (\nu + 1)  - \omega\right],  
\end{split}
\label{eq:characteristic polynomial}
\end{equation}
where we also used Eq.~\eqref{eq:stationary p0} to simplify the absolute-order term in $\omega$ (first term on the right-hand side).

The polynomial $P(\omega)$ has at least one positive root, because first, $P(0)>0$, since the absolute term is positive for $\sgn{\alpha}=1$.  Second, the coefficient in front of the cubic term in $\omega$ is negative, so that $P(\omega)\rightarrow-\infty$ for $\omega\rightarrow\infty$.
Thus, using the intermediate value theorem, it follows that $P(\omega)$ has at least one positive zero. As a consequence, any gradient-contractile system is always unstable.

Of course, this does not preclude that it could in principle be possible to stabilize gradient-contractile systems when including diffusion and/or polarity alignment in a system with a finite size.

\subsection{Scalar field only}
\label{app: scalar only}
Here we briefly discuss the limit where $\sgn{A} = 1$, $B = 0$, i.e.\ $g(p)=1$, and $\tau \rightarrow 0$ while $\beta\tau=1$. In this limit, polarity relaxation is much faster than box deformation, and Eq.~\eqref{eq:stationary p0} is solved by
\begin{equation}
  p_0=l_x^{-1}.
\end{equation}
Moreover, polarity perturbation away from this state relaxes adiabatically fast towards the scalar field perturbation. Using the linearized dynamics for the polarity, Eqs.~\eqref{eq:linearized dynamics delta px} and \eqref{eq:linearized dynamics delta py}:
\begin{align}
  \delta p_x &= ik\cos{\phi}\;\delta c\\
  \delta p_y &= ik\sin{\phi}\;\delta c.
\end{align}
Insertion in the linearized dynamics of the scalar field, Eqs.~\eqref{eq:linearized dynamics delta c}, yields:
\begin{equation}
  \bar\partial_t \delta c = \sgn{\alpha}l_x^{-2}\sin^2{\phi}\;\delta c.
\end{equation}

\subsection{Polar field only}
\label{app: polarity only}
Here, we discuss the case where $\beta = 0$, i.e.\ where the scalar field is irrelevant.  We focus on the extensile case, $\sgn{\alpha}=-1$.

\subsubsection{\texorpdfstring{Fixed magnitude, $\sgn{A} = -1, \tau\rightarrow0$}{Fixed magnitude}}
\label{app: polarity only, fixed}
In the case of fixed polarity magnitude, $\sgn{A} = -1$ and $\tau \rightarrow 0$, we have $p_0=1$ and $\delta p_x=0$. 
In this limit, Eqs.~\eqref{eq:linearized dynamics delta px} and \eqref{eq:linearized dynamics delta py} yield:
\begin{align}
  \bar\partial_t \delta p_y  &= \omega(\phi)\delta p_y \qquad\qquad\text{with} \label{eq: bpartial fixed polarity} \\    
  \omega(\phi) &= 2\nu\shr-\frac{1}{2}\Big[\nu\cos^2{(2\phi)} - \cos{2\phi}\Big].  \label{eq: omega fixed polarity only}
\end{align}
In order to plot the analytical boundaries of the phase diagram in~\autoref{fig:polar-only}j, we study the growth rates of (i) the bend mode, $\omega(\phi=0)$, (ii) the splay mode $\omega(\phi=\pi/2)$, and (iii) the maximum growth rate $\omega$ maximized over all angles, $\omega_{\mathrm{max}}$.
When at least one of the bend or the splay mode has a positive growth rate, the system is unstable. Otherwise, when both modes have negative growth rates, the system is either stable when $\omega_{\mathrm{max}}$ is negative, or stable with transiently growing modes whenever $\omega_{\mathrm{max}}$ is positive.

The growth rates of bend and splay modes are:
\begin{align}
\omega(\phi = 0) &= 2\nu\shr-\frac{1}{2}(\nu - 1) \\
\omega(\phi = \pi/2) &= 2\nu\shr-\frac{1}{2}(\nu + 1).
\end{align}
The maximum of $\omega(\phi)$ over all angles $\phi$ is: 
\begin{equation}
    \label{eq:omega max}
    \omega_{\mathrm{max}} = 
    \begin{cases}
    \omega(\phi = 0) & \text{for $\nu\leq1/2$} \\ 2\nu\shr + 1/(8\nu) & \text{for $\nu>1/2$}
    \end{cases}
\end{equation}
% Thus, for $\nu \leq 0.5$ the fastest growing mode is the bend mode, which consequently defines the phase boundaries in \autoref{fig:polar-only}j in this regime.
% Note that this strategy simplifies the analyses in the upcoming sections.
In the phase diagram in \autoref{fig:polar-only}j, there are two kinds of solid curves which define the boundaries of regions of different behavior. 
The first kind of curve, defined over the whole $\nu$ range, satisfies the $\omega(\phi = 0) = 0$ equation. This curve defines the boundaries of unstable regions as $\omega(\pi/2) < \omega(0)$ for every $(\nu, \shr)$. 
The second kind of curve, only present for $\nu >0.5$, is the $\omega_{max} = 0$ curve, which defines the region of transiently growing modes. Above this curve, both bend and splay modes have negative growth rates, but the maximal growth rate is positive. 

\begin{figure}
    \centering
    \includegraphics[width = 6.5 cm]{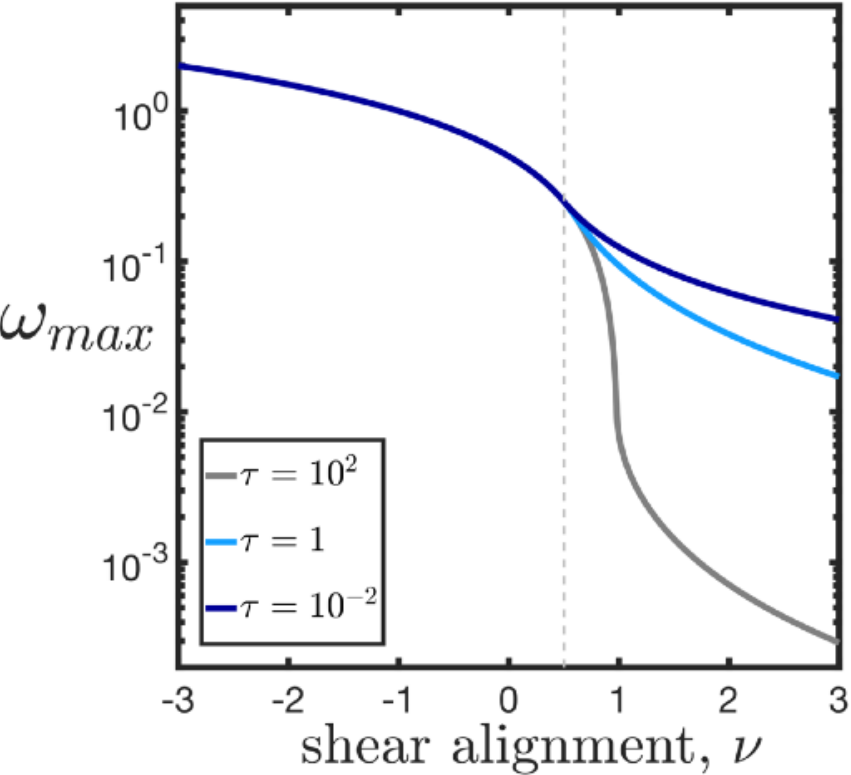}
    \caption{Maximal perturbation growth rate, $\omega_{max}$, for a polar-only system with fixed system size and finite preferred polarity magnitude. We plot three different values of polarity relaxation timescale, $\tau$. As $\tau \rightarrow +\infty$, the growth rates in the $\nu > 1$ regime converge to zero.}
    \label{fig: app fig 1}
\end{figure}

\subsubsection{\texorpdfstring{Zero preferred magnitude, $\sgn{A} = 1$}{Zero preferred magnitude}}
\label{app: polarity only, zero}
Here we derive analytical expressions for the phase boundaries in \autoref{fig:polar-only}m.
For zero preferred polarity magnitude, a finite polarity magnitude
\begin{equation}
  p_0=\sqrt{-1-\nu\shr\tau} \label{eq:polarity only, zero preferred, p0}
\end{equation}
exists only for large enough box shear rate $\vert\shr\vert > 0$, in particular $\nu\shr\tau<1$. When this condition is not met, the polarity in the stationary state has zero magnitude.
% Otherwise, there is finite $p_0$, we discuss here stability of the homogeneously deforming state.

Diagonalizing the $2\times 2$ matrix that corresponds to the dynamics of $\delta p_x$ and $\delta p_y$, Eqs.~\eqref{eq:linearized dynamics delta px} and \eqref{eq:linearized dynamics delta py}, we find the $\phi$-dependent amplitude growth rates $\omega$:
\begin{align}
  \omega_{1,2} &= \frac{1}{2} \Big[ \mathcal{T} 
  \pm \sqrt{ \mathcal{T}^2 - 4\mathcal{D} } \  \Big] , \quad\text{where}
  \label{eq:omega zero-polarity} \\
  \mathcal{T}(\phi) &= \frac{p_0^2}{2}\big(\cos{2\phi} -\nu\big) -\frac{2}{\tau}\big(1+2p_0^2\big) \\
  \mathcal{D}(\phi) &= \frac{p_0^2}{\tau}\left[ \left(\nu+\frac{4}{\tau}\right)\big(1+p_0^2\big) -\cos{2\phi}\Big(\nu \cos{2\phi} +p_0^2\Big)\right].
\end{align}
Based on Eq.~\eqref{eq:omega zero-polarity}, a mode with angle $\phi$ grows in amplitude only if $\mathcal{T}(\phi)>0$ or $\mathcal{D}(\phi)<0$.

As we argued in the section before, a system is in the unstable regime whenever the bend ($\phi=0$) or the splay ($\phi=\pi/2$) mode grows.
To see when this is the case, we first remark that $\mathcal{T}(0)>\mathcal{T}(\pi/2)$ while $\mathcal{D}(0)<\mathcal{D}(\pi/2)$. Thus, the bend mode will always be growing when the splay mode is, and so the bend mode is sufficient to decide whether the system is in the unstable regime. 
Moreover, we find that $\mathcal{D}(0)=p_0^2[-2\mathcal{T}(0) - 4p_0^2/\tau]/\tau$, from which follows that whenever $\mathcal{D}(0)\geq0$ then $\mathcal{T}(0)\leq0$.
Taken together, this means that the system is in the unstable regime if and only if $\mathcal{D}(0)<0$.
Indeed, we find that the criterion $\mathcal{D}(0)=0$ defines the boundaries of all unstable regimes in \autoref{fig:polar-only}m, where for a given parameter point $(\nu, \shr)$, the polarity magnitude $p_0$ needs to be inserted from Eq.~\eqref{eq:polarity only, zero preferred, p0}.
    
Further, if the system is not unstable, for $\mathcal{D}(0)\geq0$, the regime could be either stable or stable with transiently growing modes.  To know if there are transiently growing modes, we need to know if there is any $\phi$ with a positive $\omega$, i.e.\ with $\mathcal{T}(\phi)>0$ or $\mathcal{D}(\phi)<0$.
However, we have $\mathcal{D}(0)\geq0$ and thus $\mathcal{T}(0)\leq0$. Moreover, the maximum of $\mathcal{T}(\phi)$ is at $\phi=0$. Thus, we have $\mathcal{T}(\phi)\leq0$ for any $\phi$ outside the unstable regime.
As a consequence, any regime with transiently growing modes needs to have an angle $\phi$ for which $\mathcal{D}(\phi)<0$.
Whether this is the case depends on the sign of $\nu$.
First, for $\nu\geq 0$, the minimum of $\mathcal{D}(\phi)$ is at $\phi=0$. Thus, outside of the unstable regime, where $\mathcal{T}(0)<0$ and thus $\mathcal{D}(0)>0$, the value of $\mathcal{D}(\phi)$ can never be negative for any $\phi$. Thus, for $\nu\geq0$, there are no stable regimes with transiently growing modes (see \autoref{fig:polar-only}m).
Second, for $\nu<0$, it can be shown that the minimum of $\mathcal{D}(\phi)$ is not at $\phi=0$ only if $p_0^2<-2\nu$. In these cases, the minimal value of $\mathcal{D}(\phi)$ is
\begin{equation}
  \mathcal{D}_\mathrm{min} = \frac{p_0^2}{\tau}\left[ \left(\nu+\frac{4}{\tau}\right)\big(1+p_0^2\big) + \frac{p_0^4}{2\nu}\right].
\end{equation}
Taken together, for $\mathcal{T}(0)<0$ and $\nu<0$ the system has transiently growing modes only if $p_0^2<-2\nu$ and $\mathcal{D}_\mathrm{min}<0$.

\subsection{Scalar and polar field}
In this section we focus exclusively on the extensile case, $\sgn{\alpha}=-1$.
\subsubsection{Fixed system size, strong coupling to the scalar field}
\label{app:scalar and polar field, fixed system size}
Here we discuss the limit of strong coupling to the scalar field.
To more clearly understand what are the correct parameter regimes for strong and weak coupling to the scalar field gradient, we re-express the linearized dynamics using the following definitions:
\begin{align}
  \delta\tilde c &:= ikp_0\delta c \label{eq:tilde c}\\
  \tilde\beta &:= \frac{\beta_0}{p_0^3}=\frac{g(p_0)}{\tau p_0^2}. \label{eq:tilde beta}
\end{align}
The last equation on the second line follows directly from the stationary-state requirement for $p_0$, Eq.~\eqref{eq:stationary p0}, for fixed system size.  We now use these two definitions to simplify the linearized dynamics, Eqs.~\eqref{eq:linearized dynamics delta c}--\eqref{eq:linearized dynamics delta py}:
\begin{align}
\frac{1}{p_0^2}\bar\partial_t \delta \tilde c &= 
-\sin{\phi}\sin{2\phi} \;\delta p_x 
+ \sin{\phi}\cos{2\phi} \;\delta p_y 
\label{eq:linearized dynamics simplified delta c} \\
\frac{1}{p_0^2}\bar\partial_t \delta p_x &= 
\tilde\beta \cos{\phi} \; \delta \tilde c 
- \left[\tilde\beta G^{-1} + \frac{\nu}{2}\sin^2{2\phi}\right] \delta p_x \nonumber \\
&\qquad\qquad\qquad\quad +\frac{\nu}{2}\sin{2\phi}\cos{2\phi} \;\delta p_y 
\label{eq:linearized dynamics simplified delta px} \\
\frac{1}{p_0^2}\bar\partial_t \delta p_y &= 
\tilde\beta\sin{\phi} \;\delta \tilde c 
+ \frac{1}{2} \sin{2\phi}\big(\nu \cos{2\phi}-1\big) \;\delta p_x \nonumber \\
&\qquad -\left[\tilde\beta + \frac{1}{2}\cos{2\phi}\big(\nu \cos{2\phi}-1\big) \right]\delta p_y.  \label{eq:linearized dynamics simplified delta py}
\end{align}
In Eq.~\eqref{eq:linearized dynamics simplified delta px}, we have also used the definition of $G(p_0)$, Eq.~\eqref{eq:G}.
From Eqs.~\eqref{eq:linearized dynamics simplified delta px} and \eqref{eq:linearized dynamics simplified delta py}, we see directly that $\tilde\beta$ compares the scalar field coupling strength (and the polarity magnitude control) with the flow-induced feedback.
Thus, the correct limit for a strong scalar field coupling is $\tilde\beta\gg1$, i.e.\ $\beta\gg p_0^3$.

For strong scalar field coupling, $\tilde\beta\gg1$, 
Eqs.~\eqref{eq:linearized dynamics simplified delta px} and \eqref{eq:linearized dynamics simplified delta py} imply that polarity relaxes adiabatically fast towards:
\begin{align}
  \delta p_x &= iGp_0k_x\delta c \label{eq:app large-beta_delta px}\\
  \delta p_y &= ip_0k_y\delta c \label{eq:app large-beta_delta py}.
\end{align}
These equations can be used to obtain a criterion for the stability of the system. Indeed, combining Eqs.~\eqref{eq: bpartial delta c}, \eqref{eq:co-deforming velocity}, \eqref{eq:active stress-polarity perturbation}, \eqref{eq:app large-beta_delta px}, and \eqref{eq:app large-beta_delta py}, we obtain:
\begin{equation}
  \partial_t\delta c = p_0^2\,\sin^2{\phi}\,\Big[2(1-G)\cos^2{\phi} - 1\Big]\,\delta c.
\end{equation}
From this equation directly follows that the system is marginally stable whenever $G\geq1/2$.

Alternatively, to allow for the more intuitive explanation in the main text, we introduce an angle $\theta_p$, which globally characterizes the orientation of $\bm{\delta p}$.  However, for a given perturbation mode with wave vector $\bm{k}$, the direction of $\bm{\delta p}$ will spatially depend on the phase of the Fourier mode. We remove this ambiguity by dividing $\bm{\delta p}$ by $ik\delta c$, defining the angle $\theta_p$ by
\begin{equation}
  \frac{\bm{\delta p}}{ik\delta c} = \hat{p}\begin{pmatrix} \cos{\theta_p} \\ \sin{\theta_p} \end{pmatrix},\label{eq:definition theta_p}
\end{equation}
where $\hat{p}>0$.
From Eqs.~\eqref{eq:app large-beta_delta px}, \eqref{eq:app large-beta_delta py}, and \eqref{eq:definition theta_p} then follows for the norm $\hat{p} = p_0\hat{G}$, where $\hat{G}=[G^2\cos^2{\phi}+\sin^2{\phi}]^{1/2}$, and for the angle:
\begin{equation}
  \tan{\theta_p}=\frac{1}{G}\tan{\phi}.
\end{equation}
This is Eq.~\eqref{eq:theta_p} in the main text.
We proceed similarly to define the angle of the active stress perturbation nematic, $\theta_\sigma$. More precisely, we define $\theta_\sigma$ as the angle of the nematic $-\delta\tilde\sigma^a_{ij}/(ik\delta c)$:
\begin{equation}
  -\frac{\bm{\delta\tilde\sigma^a}}{ik\delta c} = \hat{\sigma}\begin{pmatrix} 
  \cos{2\theta_\sigma} & \sin{2\theta_\sigma} \\
  \sin{2\theta_\sigma} & -\cos{2\theta_\sigma}
  \end{pmatrix},\label{eq:definition theta_sigma}
\end{equation}
where $\hat{\sigma}>0$.
Together with Eqs.~\eqref{eq:active stress-polarity perturbation} and \eqref{eq:definition theta_p}, we have indeed
\begin{equation}
  \theta_\sigma = \frac{\theta_p}{2}.
\end{equation}
and $\hat\sigma = -\alpha p_0\hat{p} = -\alpha p_0^2\hat{G}$.
Insertion of Eq.~\eqref{eq:definition theta_sigma} into the equation for the velocity perturbation $\delta\bar\v_x$, Eq.~\eqref{eq:co-deforming velocity}, and noting that for fixed system size $\delta\bar\v_x=\delta\v_x$, yields:
\begin{equation}
  \delta\v_x = - p_0^2\hat{G}\,\sin{\phi}\,\sin{\big(2[\theta_\sigma-\phi]\big)}\,\delta c.
\end{equation}
Together with Eq.~\eqref{eq: bpartial delta c}, this results in 
\begin{equation}
  \partial_t\delta c = p_0^2\hat{G}\,\sin{\phi}\,\sin{\big(2[\theta_\sigma-\phi]\big)}\,\delta c.
\end{equation}
Hence, the system is unstable whenever there is a $\phi$ with positive $\sin{\phi}\,\sin{\big(2[\theta_\sigma-\phi]\big)}$.

\subsubsection{Deforming system}
\label{app: analytical lines}
In this part we derive the analytical curves that define the unstable regions in \autoref{fig:scalar and polar, deforming}. As discussed in \autoref{app: polarity only}, the system is in the unstable regime, whenever a mode grow with $\phi=0$ or with $\phi=\pi/2$ grows.

For $\phi=0$, the linearized dynamics of $\delta p_y$, \eqref{eq:linearized dynamics delta py}, decouple from those of $\delta c$ and $\delta p_x$, Eqs.~\eqref{eq:linearized dynamics delta c} and \eqref{eq:linearized dynamics delta px}. We find for the growth rate of the orientational perturbations, $\delta p_y$:
\begin{equation}
  \omega(\phi=0) = - \frac{\beta_0}{p_0^2} + 2 \nu\shr -\frac{p_0^2}{2}(\nu - 1).
\end{equation}
Moreover, for $\phi=\pi/2$, the linearized dynamics of $\delta c$ and $\delta p_y$ decouples from that of $\delta p_x$, where it can be directly shown that the maximum growth rate of the $\phi=\pi/2$ modes is always smaller than $\omega(\phi=0)$. Thus, the $\phi=0$ growth rate decides whether the system is in the unstable regime or not.  Hence, all boundaries to the unstable regime are given by $\omega(\phi=0)=0$ (all black solid curves in \autoref{fig:scalar and polar, deforming}). To obtain a relation between $\nu$ and $\shr$, for the fixed norm case, $p_0$ was set to one (\autoref{fig:scalar and polar, deforming}a-c), while for ZPM polarity, the polarity magnitude $p_0$ was eliminated using Eq.~\eqref{eq:p0} (\autoref{fig:scalar and polar, deforming}d-f).

\section{System size effects}
\label{app:system-size effects}
Here we report analytical results for the minimal diffusion term, $D_\mathrm{crit}k^2$ required to stabilize the system, depending on the parameters $\beta$ and $\nu$, the magnitude of the polarity alignment term $\kappa k^2$, and the sign of the activity $\mathrm{sgn}(\alpha)$.
We focus on a non-deforming system with fixed polarity magnitude.  As discussed in the main text, without diffusion or polarity alignment the system is always unstable in this case.

The linearized perturbation dynamics read
\begin{align}
\bar\partial_t \delta c &= -Dk^2\delta c
+ \frac{i\sgn{\alpha}}{k}\sin{\phi}\cos{2\phi} \;\delta p_y \\
\bar\partial_t \delta p_y &= 
i\beta k \sin{\phi} \;\delta c + \bigg[\frac{\sgn{\alpha}}{2}\cos{2\phi}\big(\nu \cos{2\phi}-1\big) \nonumber\\
&\qquad\qquad\qquad\qquad\qquad-\beta - \kappa k^2\bigg]\delta p_y.
\end{align}
The perturbation growth rates $\omega_{1,2}(\phi)$ are the eigen values of the coefficient matrix.  Denoting trace and determinant of this matrix by $\mathcal{T}(\phi)$ and $\mathcal{U}(\phi)$, respectively, we have:
\begin{equation}
\omega_{1,2}(\phi) = \frac{1}{2}\left[\mathcal{T}(\phi) \pm \sqrt{\mathcal{T}^2(\phi) - 4\mathcal{U}(\phi)}\right].
\end{equation}
The system is stable only iff both $\omega_1$ and $\omega_2$ are negative, i.e.\ iff $\mathcal{T}<0$ and $\mathcal{U}>0$, for all angles $\phi$.

As a consequence, we find after some calculation that for gradient-extensile activity, $\mathrm{sgn}(\alpha)=-1$, the system can be stabilized by diffusion iff
\begin{equation}
  \beta+\kappa k^2 >
    \begin{cases}
      (1-\nu)/2 & \text{for $\nu\leq1/2$ and} \\
      1/8\nu & \text{for $\nu>1/2$.}
    \end{cases}
\end{equation}
In both cases, the critical value of the diffusion term is given by
\begin{equation}
\begin{split}
  D_\mathrm{crit}k^2 &= \beta\Bigg[-1+4(\beta+\kappa k^2) \\
  &\quad\qquad + 4\sqrt{(\beta+\kappa k^2)\Big(\beta+\kappa k^2-[1-\nu]/2\Big)}\Bigg]^{-1}.
\end{split}
\end{equation}

For gradient-contractile activity, $\mathrm{sgn}(\alpha)=1$, the system can be stabilized by diffusion iff 
\begin{equation}
  \beta+\kappa k^2 >
    \begin{cases}
      (1+\nu)/2 & \text{for $\nu\geq-1/2$ and} \\
      -1/8\nu & \text{for $\nu<-1/2$.}
    \end{cases}
\end{equation}
For $\nu\geq-1/2$ or whenever $\nu<-1/2$ and $\beta+\kappa k^2>(1-\nu)/6$ the critical value of the diffusion term is
\begin{equation}
  D_\mathrm{crit}k^2 = \frac{\beta}{\beta+\kappa k^2 -[1+\nu]/2}.
\end{equation}
Meanwhile, whenever $\nu<-1/2$ and $(1-\nu)/6>\beta+\kappa k^2>-1/8\nu$ the critical value of the diffusion term is 
\begin{equation}
\begin{split}
  D_\mathrm{crit}k^2 &= \beta\Bigg[-1-4(\beta+\kappa k^2) \\
   &\quad\qquad+ 4\sqrt{(\beta+\kappa k^2)\Big(\beta+\kappa k^2+[1-\nu]/2\Big)}\Bigg]^{-1}.
\end{split}
\end{equation}

\bibliography{Oriented-Deformation-Theory}% Produces the bibliography via BibTeX.

\end{document}